\documentclass[twocolumn,prd, aps,superscriptaddress,preprintnumbers,tightenlines,showpacs,nofootinbib,eqsecnum,amsfonts,amsmath]{revtex4}

\usepackage{epsfig}
\usepackage{graphics}
\usepackage{graphicx}
\usepackage{amsmath,amssymb,mathrsfs}
\usepackage{amsfonts}
\usepackage{color}
\usepackage{wasysym}
\usepackage{times}
\usepackage{mathptmx}
\usepackage{gensymb}

\begin{document}

\newcommand{\be}{\begin{equation}}
\newcommand{\ee}{\end{equation}}
\newcommand{\ber}{\begin{eqnarray}}
\newcommand{\eer}{\end{eqnarray}}
\newcommand{\bea}{\begin{eqnarray}}
\newcommand{\eea}{\end{eqnarray}}
\newcommand{\ie}{i.e.}
\newcommand{\dt}{{\rm d}t}
\newcommand{\df}{{\rm d}f}
\newcommand{\dtheta}{{\rm d}\theta}
\newcommand{\dphi}{{\rm d}\phi}
\newcommand{\rhat}{\hat{r}}
\newcommand{\iotahat}{\hat{\iota}}
\newcommand{\phihat}{\hat{\phi}}
\newcommand{\hc}{{\sf h}}
\newcommand{\etal}{\textit{et al.}}
\newcommand{\balpha}{{\bm \alpha}}
\newcommand{\bbeta}{{\bm \psi}}
\newcommand{\fmerg}{f_{1}}
\newcommand{\fring}{f_{2}}
\newcommand{\fcut}{f_{3}}
\newcommand{\rmi}{{\rm i}}
\def\rd{{\textrm{\mbox{\tiny{RD}}}}}
\def\qnr{{\textrm{\mbox{\tiny{QNR}}}}}
\newcommand{\A}{\mathcal{A}}
\newcommand{\NS}{\mathrm{NS}}
\newcommand{\CC}{\mathcal{C}}

\newcommand{\ajith}[1]{\textcolor{green}{\textit{Ajith: #1}}}
\newcommand{\sascha}[1]{\textcolor{blue}{\textit{Sascha: #1}}}
\newcommand{\mdh}[1]{\textcolor{red}{\textit{Mark: #1}}}
\newcommand{\patricia}[1]{\textcolor{magenta}{\textit{Patricia: #1}}}

\newcommand{\LIGOCaltech}{LIGO Laboratory, California Institute of Technology, 
Pasadena, CA 91125, USA}
\newcommand{\TAPIR}{Theoretical Astrophysics, California Institute of Technology, 
Pasadena, CA 91125, USA}
\newcommand{\UIB}{Departament de F\'isica, Universitat de les Illes Balears, 
Crta. Valldemossa km 7.5, E-07122 Palma, Spain}
\newcommand{\Vienna}{Gravitational Physics, Faculty of Physics, University of Vienna, Boltzmanngasse
5, A-1090 Vienna, Austria}
\newcommand{\Cardiff}{School of Physics and Astronomy, Cardiff University, Queens Building, CF24 3AA, Cardiff, United Kingdom}

\newtheorem{definition}{Definition}
\newtheorem{theorem}{Theorem}
\newtheorem{proposition}{Proposition}
\newtheorem{lemma}{Lemma}
\newtheorem{observation}{Observation}



\title{Tracking the precession of compact binaries from their gravitational-wave signal}

\author{Patricia Schmidt}
\affiliation{\Cardiff}
\affiliation{\Vienna}

\author{Mark Hannam}
\affiliation{\Cardiff}
\affiliation{\Vienna}

\author{Sascha Husa}
\affiliation{\UIB}

\author{P.~Ajith}
\affiliation{\LIGOCaltech}

\begin{abstract}
We present a simple method to track the precession of a black-hole-binary system, 
using only information from the gravitational-wave (GW) signal.
Our method consists of locating the frame from which the magnitude of the $(\ell=2,|m|=2)$ 
modes is maximized, which we denote the ``quadrupole-aligned'' frame. 
We demonstrate the efficacy of this method when applied to waveforms
from numerical simulations. In the test case of an equal-mass nonspinning binary, our method
locates the direction of the orbital angular momentum to within 
$(\Delta \theta, \Delta \varphi) = (0.05^{\circ},0.2^{\circ})$. 
We then apply the method to a $q = M_2/M_1 = 3$ binary that exhibits significant precession. 
In general a spinning binary's 
orbital angular momentum $\mathbf{L}$ is \emph{not} orthogonal to the orbital plane. 
Evidence that our method locates the direction of $\mathbf{L}$ rather than the normal of 
the orbital plane is provided by comparison with post-Newtonian (PN) results. 
Also, we observe that it accurately reproduces similar higher-mode amplitudes to a 
comparable non-spinning (and therefore non-precessing) 
binary, and that the frequency of the $(\ell=2,|m|=2)$ modes is consistent with the 
``total frequency''  of the binary's motion. 
The simple form of the quadrupole-aligned waveform will be useful in attempts
to analytically model the inspiral-merger-ringdown (IMR) signal of precessing binaries, and in
standardizing the representation of waveforms for studies of accuracy and consistency of 
source modelling efforts, both numerical and analytical.
\end{abstract}

\maketitle

\section{Introduction}
\label{sec:intro}

Black-hole-binary mergers are expected to be key sources for gravitational-wave
(GW) astronomy~\cite{lrr-2009-2}. Accurate theoretical models of the 
GW signal are necessary to both detect these sources and to determine their physical 
parameters and their location in the universe. The GW signal from the inspiral can be 
calculated by analytic approximation techniques~\cite{lrr-2006-4,Damour:2009ic}, 
and the merger of the two black 
holes and ringdown of the final black hole can be calculated from numerical simulations
in full General 
Relativity~\cite{Pretorius:2005gq,Campanelli:2005dd,Baker:2005vv,Centrella:2010mx}.

Numerical simulations can produce waveforms for only discrete points in the 
parameter space of binary configurations, but significant progress has been made
in synthesizing information from post-Newtonian (PN) and effective-one-body (EOB) 
methods, numerical relativity (NR), 
and perturbation theory, to produce analytic models of the complete 
inspiral-merger-ringdown signal over some regions of the parameter 
space. Most models to date treat nonspinning 
binaries~\cite{Ajith:2007qp,Ajith:2007kx,Ajith:2007xh,Buonanno:2007pf,Damour:2007yf,Damour:2007vq,Damour:2008te,Baker:2008mj,Mroue:2008fu,Damour:2009kr,Buonanno:2009qa}, or binaries in
which the black-hole spins do not precess~\cite{Ajith:2009bn,Santamaria:2010yb,Pan:2009wj} 
(although there has been one 
first attempt at a precession model~\cite{Sturani:2010yv}). 

Precession adds a number of complications.
When the spins are not parallel to the orbital angular momentum 
their orientation varies with time, as does the orbital angular momentum itself; the
orbital plane precesses. Both the precession of the spins and of the orbital 
plane each introduce modulations into the GW amplitude, oscillations into the GW 
frequency, and variations in the distribution of signal power across different
harmonics of the waveform. All of these complicate efforts to produce an 
analytic model of precessing-binary waveforms. In addition, they make it difficult
to uniquely characterize the wave signal. For example, the total phase of the 
dominant mode of the signal depends on the initial orientation of the orbital 
plane. This makes it difficult to determine if two waveforms were produced
by the same binary configuration, or to compare independent numerical simulations,
a task that is relatively simple for non-precessing non-eccentric 
binaries~\cite{Baker:2007fb,Pan:2007nw,Hannam:2009hh}.

We propose a method to put a precessing-binary waveform into a particularly simple
form. The method is based on finding a preferred time dependent 
coordinate system for the gravitational wave signal, which tracks the precession.

Gravitational wave signals are most conveniently expressed in terms of
spherical harmonics of spin-weight $s=-2$, $Y^s_{lm}(\theta,\varphi)$, where
$(\theta,\varphi)$ are the standard polar coordinates on the unit sphere. 
The dominant modes are the quadrupole modes, where $\ell=2$, and $-2 \leq m \leq 2$.
If the system is rotated,  the modes of a particular
$\ell$ mix among each other according to the transformation law described in
Appendix \ref{sec:appendix}.

As can be seen from standard post-Newtonian and post-Minkowskian descriptions,
binary systems emit gravitational waves predominantly in the
direction orthogonal to the orbital plane. Correspondingly, if our system is oriented 
such that this direction is along the $z$-axis, 
then we expect that the dominant signal is given by the $(\ell=2,\vert m \vert =2)$ spherical harmonics of the wave. The modes $\vert m \vert = 1$  vanish 
when the two black holes can be exchanged by symmetry, and $m=0$ is a 
non-oscillating mode related to memory effects, see e.g.,~\cite{Favata:2009ii,Pollney:2010hs}.
If we choose different (rotated) coordinates $(\theta',\varphi')$ to define 
a new basis $Y^s_{lm}(\theta',\varphi')$, 
then mode mixing will complicate the 
spherical harmonic description of the signal, and for example even an equal mass nonspinning
binary will exhibit nonvanishing $\vert m \vert = 1$  modes. We illustrate this effect in
Sec.~\ref{sec:emns}.

Therefore, we can determine a preferred direction \emph{from the wave 
signal alone} by finding the orientation that maximizes the $(\ell=2,|m|=2)$ modes. 
This is the method that we will discuss in this paper, and we will refer to
waveforms that are given in terms of spherical harmonics that
are aligned with this direction as
``quadrupole-aligned'' waveforms.

In a precessing system there are two contributions to the frequency of the binary
motion: the frequency of the motion about the orbital-plane axis, $\omega_{\rm orb}$,
which will increase during a non-eccentric inspiral as a monotonic function, 
and the frequency of the motion due to the precessional motion,
which will oscillate as a function of time. 
The total frequency of the motion is $\omega = \omega_{\rm orb} - \dot{\varphi} 
\cos \theta$, where $\theta$ is the inclination of the normal to the orbital plane 
from the $z$-axis, and $\varphi$ is the rotation of the normal about the $z$-axis
in the $xy$ plane.
(This corresponds to the result in Eq.~(3.10) in~\cite{Arun:2008kb}.)
In a kinematical description of the binary, these two frequencies together prescribe the
bodies' acceleration, which is the dominant source of gravitational radiation. One of 
the properties we expect from our quadrupole-aligned waveform is that during the 
inspiral the frequency of the $(\ell=2,|m|=2)$ modes will to a good approximation 
satisfy the relation \begin{equation}
\omega_{22} = 2(\omega_{\rm orb} - \dot\varphi \cos \theta). 
\label{eqn:freqreln}
\end{equation}

Our main results are that (1) we can determine the quadrupole-aligned direction from 
the GW signal to high accuracy (within a fraction of a degree during most of the inspiral), and
(2) the GW signal is indeed far simplified, see in particular Fig.~\ref{fig:frequency} of the GW 
frequency before and after our (2,2)-maximization procedure, where the final frequency 
does indeed satisfy Eq.~(\ref{eqn:freqreln}).
In addition, we show that the GW signal 
is emitted in the direction of the orbital angular momentum of the binary, which
is \emph{not} in general perpendicular to the orbital plane. We illustrate this
effect with an example from PN theory, where it can be seen explicitly that 
the effective orbital angular momentum is not parallel to the naive Newtonian
angular momentum. 

In Sec.~\ref{sec:algorithm} we provide details of our algorithm to find the
orbital-angular-momentum direction from the GW signal, and in Sec.~\ref{sec:numrel}
describe our numerical methods and numerical simulations. The results of our method
are presented in Sec.~\ref{sec:results}, where we verify our method using a simple test
case of an equal-mass nonspinning binary, and then apply the method to an unequal-mass
spinning binary that undergoes significant precession. We discuss these results and 
prospects for future work in Sec.~\ref{sec:conclusion}.

\section{Maximization procedure algorithm} 
\label{sec:algorithm}

The Weyl scalar $\Psi_4$ as calculated from the numerical code is decomposed 
into standard spin-weighted spherical harmonics (see \cite{Brugmann:2008zz}
for our implementation). If the orbital angular momentum of the binary is parallel to the 
$z$-axis, then the GW signal will be dominated by the $(\ell=2,|m|=2)$ modes.
We also expect that the coefficient of the $(\ell=2,|m|=2)$ modes
will be maximal in this case; for any other orientation of the orbital angular 
momentum, the $(\ell=2,|m|=2)$ modes will be weaker. 

Given the $\ell=2$ modes $\Psi'_{4,2m}$ from the numerical code, we can rotate 
the frame to any other orientation using the transformation described in 
Appendix~\ref{sec:appendix}, to produce the corresponding $\Psi_{4,2m}$ in that frame. 
We locate the direction of the orbital angular
momentum by searching over a 
range of the Euler angles $(\beta,\gamma)$ to find a global maximum in 
$\Psi_{4,22}$. 

The procedure in practice is as follows. We start our analysis after the passage of
the pulse of junk radiation. Since we extract the GW signal at either 
$R_{ex} = 90M$ or $R_{ex} = 100M$,
we take the start time to be at about $t = 150M$. At this time, we produce a first guess of the direction
of $\mathbf{L}$ from the location of the black-hole punctures at that time.
This provides a guess $(\beta_0,\gamma_0)$ of the Euler angles by which 
to rotate the system. 
Given this initial guess, we then search over a range of 
$(\beta,\gamma) = (\beta_0 \pm 10^{\circ}, \gamma_0 \pm 10^{\circ})$ 
with an angular
resolution of $0.1^{\circ}$, and find the angle for which the function 
$|\Psi_{4,22}|^2 + |\Psi_{4,2-2}|^2$ has a maximum. In our test cases, 
where the orientation is constant, this procedure is trivial, but in general this first guess 
may not be very accurate. In particular, it does not take into account the time lag from the 
source to the GW extraction sphere. However, we do not expect the system to 
precess by as much as $10^{\circ}$ over $\approx 100M$ of evolution. We also know that
the Newtonian orbital angular momentum $\mathbf{L_N}$ calculated from the 
puncture motion is not in general parallel to the direction that maximizes the 
$(\ell=2,m=2)$ mode, but we do not expect the deviation to be larger than a 
few degrees; we will discuss this point further at the end of Sec.~\ref{sec:results}.

For subsequent times, we use the angles from the previous time step 
as the first guess, and now search over the smaller range of $\pm3^{\circ}$ in
each angle. We locate the maximum in $|\Psi_{4,22}|^2 + |\Psi_{4,2-2}|^2$ with a quadratic
curve fit through the data from the search. 

\begin{figure}[t]
\centering
\includegraphics[width=80mm]{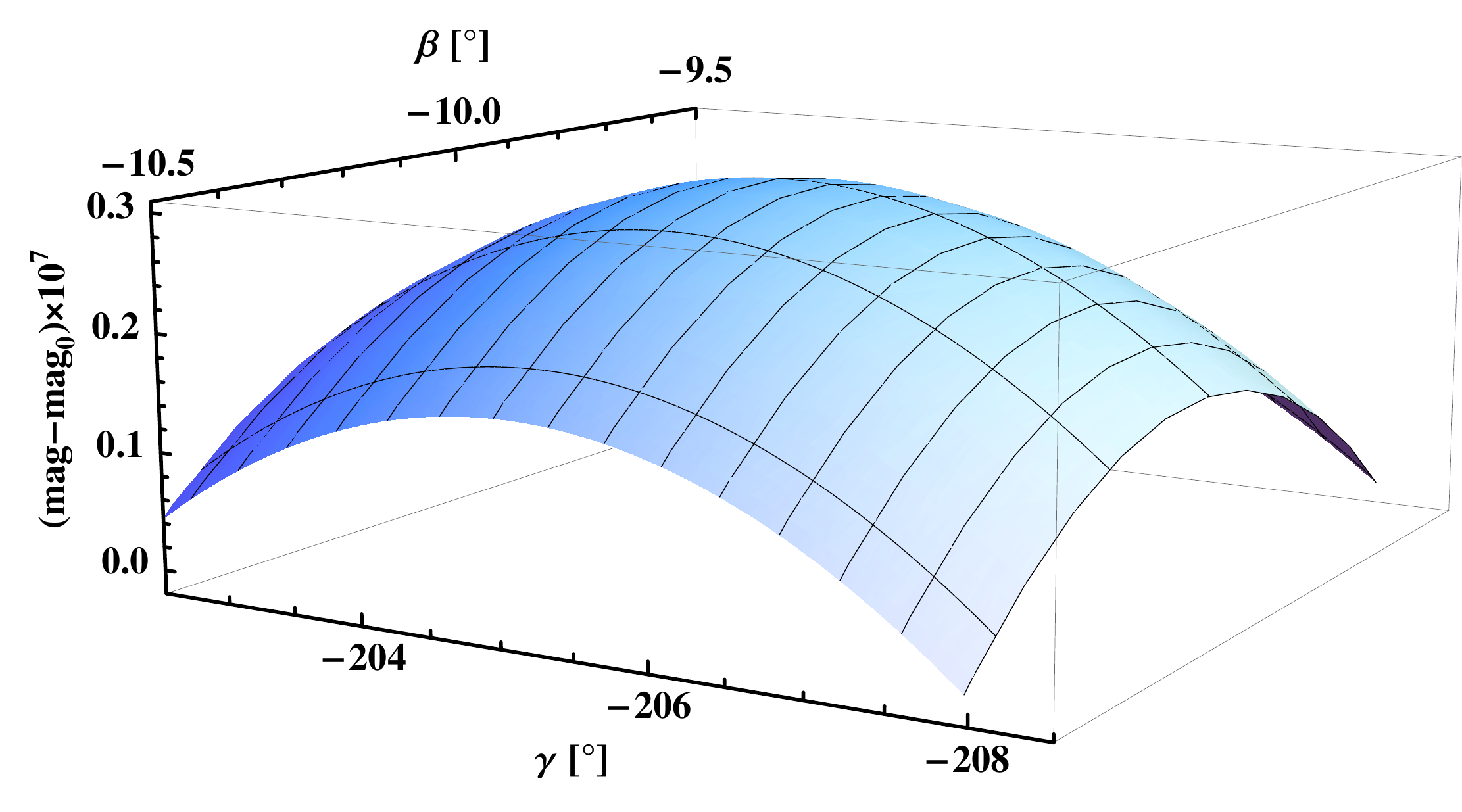}
\caption{
Profile of the magnitude of $\Psi_{4,22}$ as the system is rotated by
the Euler angles $\beta$ and $\gamma$. The example is taken from 
one time step ($t = 562M$) of the rotated equal-mass nonspinning 
case discussed in Sec.~\ref{sec:emns}. Note that there is a clearly
defined maximum, which in this case is at $(\beta,\gamma) = (-10^{\circ},-205^{\circ})$.
}
\label{fig:maximization}
\end{figure}

At all times we find a clear maximum in the amplitude of $\Psi_{4,22}$ as a 
function of the rotation angles. An example is given in 
Fig.~\ref{fig:maximization}, based on one time step of the rotated 
equal-mass nonspinning case presented in Sec.~\ref{sec:emns}.

\section{Numerical methods and Simulations} 
\label{sec:numrel}

We performed numerical simulations with the BAM code
\cite{Brugmann:2008zz,Husa:2007hp}. 
The code starts with black-hole-binary puncture initial data 
\cite{Brandt:1997tf,Bowen:1980yu} generated using a pseudo-spectral 
elliptic solver~\cite{Ansorg:2004ds}, and evolves them with the 
$\chi$-variant of the
moving-puncture \cite{Campanelli:2005dd,Baker:2005vv,Hannam:2006vv}
version of the BSSN
\cite{Shibata:1995we,Baumgarte:1998te} formulation of the 3+1 Einstein 
evolution equations. Spatial finite-difference derivatives are
sixth-order accurate in the bulk \cite{Husa:2007hp}, Kreiss-Oliger
dissipation terms converge at fifth order, and a fourth-order Runge-Kutta
algorithm is used for the time evolution. 
The gravitational waves emitted by the binary are calculated from the
Newman-Penrose scalar $\Psi_4$, and the details of our implementation of
this procedure are given in \cite{Brugmann:2008zz}. 
See e.g.~\cite{Hannam:2010ec} for a recent extensive parameter study of non-precessing binaries that uses the same numerical code and general setup.

In each simulation, the black-hole punctures are initially a coordinate distance
$D$ apart, and are placed on the $y$-axis at $y_1 = -qD/(1+q)$ and $y_2 = D/(1+q)$,
where $q = M_2/M_1$ is the ratio of the black hole masses in the binary, and we always choose 
$M_1 < M_2$. The masses $M_i$ are estimated from the Arnowitt-Deser-Misner
(ADM) mass at each puncture, according to the method described 
in~\cite{Brandt:1997tf}; see also the Appendix of~\cite{Hannam:2010ec}. 
The Bowen-York punctures are given 
momenta $p_x = \mp p_t$ tangential to their separation vector, and $p_y = \pm p_r$
towards each other. The latter momentum component accounts for the (initially small) radial 
motion of the black holes as they spiral together. Initial parameters for low-eccentricity 
inspiral were produced using integrations of the PN equations of motion, 
as described in~\cite{Husa:2007rh,Hannam:2010ec}. 

The eccentricity is measured with respect to the frequency of the orbital motion, as
in all of our past work on eccentricity 
removal~\cite{Husa:2007rh,Gopakumar:2007vh,Hannam:2007wf,Hannam:2010ec}, and
also discussed in~\cite{Mroue:2010re,Buonanno:2010yk} and references therein. The
eccentricity is estimated as the extrema of 
$e_\omega(t) = (\omega(t) - \omega_{QC}(t))/(2 \omega_{QC}(t))$, where $\omega$ is the 
frequency of the $(\ell=2,m=2)$ mode of the waveform, and $\omega_{QC}(t)$ 
is an estimate of the frequency evolution for a non-eccentric binary, calculated by a 
smooth curve fit through the numerical data. 

The grid setup is similar to that used in~\cite{Brugmann:2008zz}, and using
the notation introduced there, the simulations discussed in this paper all use a
configuration of the form
$\chi_{M\eta=2}[l_1\times N:l_2\times 2N:6]$.  This indicates that the 
simulation used the $\chi$ variant of the moving-puncture method, $l_1$ nested
mesh-refinement boxes with a base value of $N^3$ points surround each black hole,
and $l_2$ nested boxes with $(2N)^3$ points surround the entire system, and there are
six mesh-refinement buffer points. The $\eta$ parameter in the BSSN system is $M\eta = 2$.
The choices of $N$, $l_1$, $l_2$ and the resolutions
are given in Tab.~\ref{tab:parameters}. The resolution around the puncture is denoted by 
$M_1/h_{min}$, which is the resolution with respect to the \emph{smallest black hole}, 
$M_1$. The puncture of the second black hole will have the same numerical resolution, 
but if the black hole is bigger, $M_2>M_1$, then it will effectively be better resolved. 
In unequal-mass cases, different numbers of refinement levels can be used around each
black hole, so that the larger black hole need not be unnecessarily well-resolved, which
would slow down the code.

Far from the sources, the meaningful length scale is
the total mass of the binary, $M = M_1+M_2$, and so the resolution on the coarsest level is
given by $h_{max}/M$. 

We consider two configurations. The first is an equal-mass nonspinning binary, 
using the same setup as first described in~\cite{Hannam:2007ik}. The 
initial separation is $D=12M$, and the binary completes about nine orbits before
merger. One additional simulation was performed in which the orbital plane 
was first rotated by $10^{\circ}$ about the $y$-axis, and then around the $z$-axis by 
$25^{\circ}$. 

The second configuration is a binary with mass ratio $q=3$, where the larger black 
hole has spin $S_2/M_2^2 = 0.75$. In the calculation of the initial parameters, the spin
is directed perpendicular to the orbital angular momentum when the binary is at
a separation of $D=30M$. The configuration is evolved using the PN equations of
motion until about $D=10M$, and the momenta read off from the PN evolution at
a point where the point particles pass through the $xy$ plane. This leads to the 
parameters given in Tab.~\ref{tab:parameters}. 

Some key physical properties of the simulations are given in the last three rows of
Tab.~\ref{tab:parameters}: the estimate of the eccentricity of the binary, the time when
the GW signal reaches its peak amplitude, and the number of GW cycles up until that
time.

\begin{table}[htp]
\caption{\label{tab:parameters}
Parameters for the two configurations that we consider in this paper, the equal-mass
nonspinning case, and the $q=3$ precessing-spin case. (For the rotated equal-mass
nonspinning case, the momenta are 
$\mathbf{p}_i = \mp\{0.07567,0.03588,0.01477\}$.) 
The lower rows of the table indicate
the numerical grid, which follows the conventions of~\cite{Brugmann:2008zz,Hannam:2010ec}.
}
\begin{ruledtabular}
\begin{tabular}{||c || c | c ||}
 $q$  & 1 & 3 \\
 $m_i$ & $\{0.488278,0.488278\}$ & $\{0.4779361,1.0234487\}$ \\
 $\mathbf{S}_1$ & $\{0,0,0\}$ & $\{0,0,0\}$  \\
 $\mathbf{S}_2$ & $\{0,0,0\}$ & $\{-1.048, 1.199, 0.560\}$  \\
 $\mathbf{x}_1$ & $\{0,6,0\}$ & $\{0, 15.0779, 0\}$ \\
 $\mathbf{x}_2$ & $\{0,-6, 0\}$ & $\{0, -5.02598, 0\}$ \\
 $D/M$                 & $12.00$      & $10.05$ \\
 $\mathbf{p}_x$ & $\mp 0.085035$  &  $\mp0.126292$ \\
 $\mathbf{p}_y$ & $\pm0.000537$   &  $\mp0.00139578$ \\
 $\mathbf{p}_z$ & $0$                         &  $\pm0.0696932$ \\
 \hline
 $N$                            & 64      & 96 \\
 $(l_1,l_2)$                & (5,5)  & (4/5,8)  \\
 $M_1/h_{\rm min}$ & 21.3 & 60.0 \\
 $h_{\rm max}/M$     & 12 & 17.06 \\
 $x_{i,{\rm max}}/M$ & 774  & 1647 \\
 \hline
 e &  0.0016  &  0.0015 \\
 $t_{\rm peak}/M$ & 1940 & 1170 \\
 $N_{\rm cycles}$ & 19 & 14
\end{tabular}
\end{ruledtabular}
\end{table}

\section{Numerical results}
\label{sec:results}

\subsection{Test case: equal-mass nonspinning binary}
\label{sec:emns}

In order to test our maximization procedure, we consider two simulations of 
an equal-mass nonspinning binary. In one, a reference case, the orbital angular momentum 
is oriented parallel to the $z$-axis, and so the waveform is already in the quadrupole-aligned
frame. The simulation starts at $D=12M$ and covers about nine
orbits before merger. 

\begin{figure}[t]
\centering
\includegraphics[width=80mm]{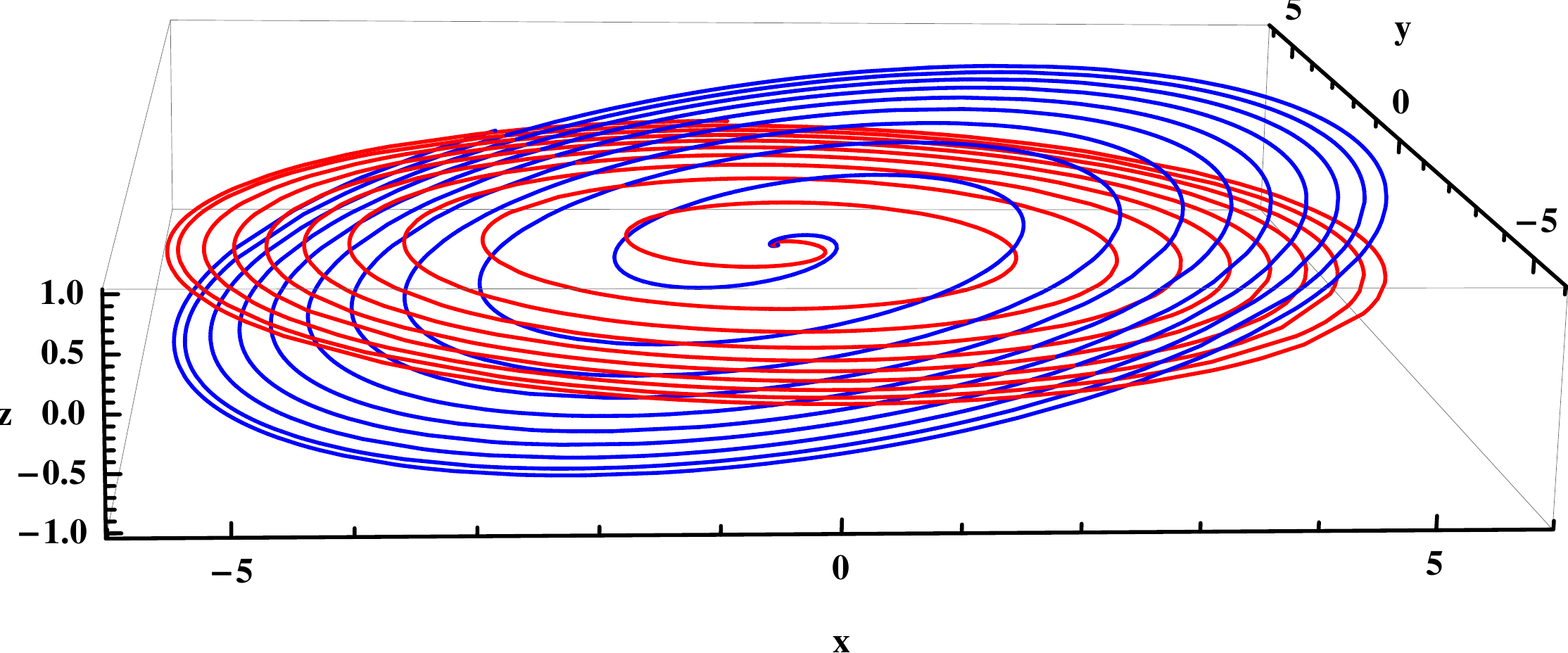}
\caption{
Motion of one black-hole puncture for the reference and rotated cases. 
The orbital planes are related by a rotation about the $y$-axis of $10^{\circ}$,
and about the $z$-axis of $25^{\circ}$.
}
\label{fig:twistmotion}
\end{figure}

\begin{figure*}[t]
\centering
\includegraphics[width=80mm]{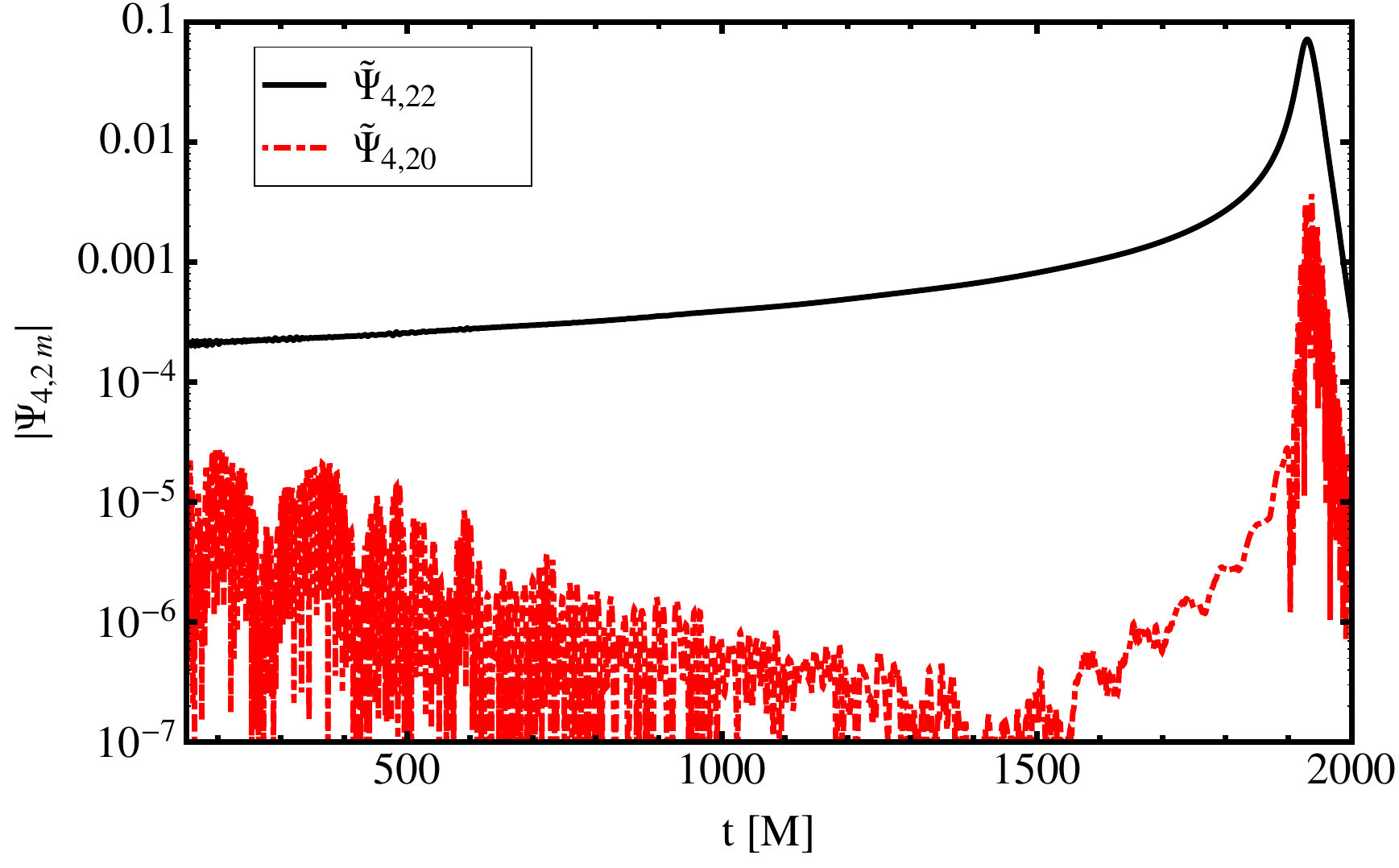}
\includegraphics[width=80mm]{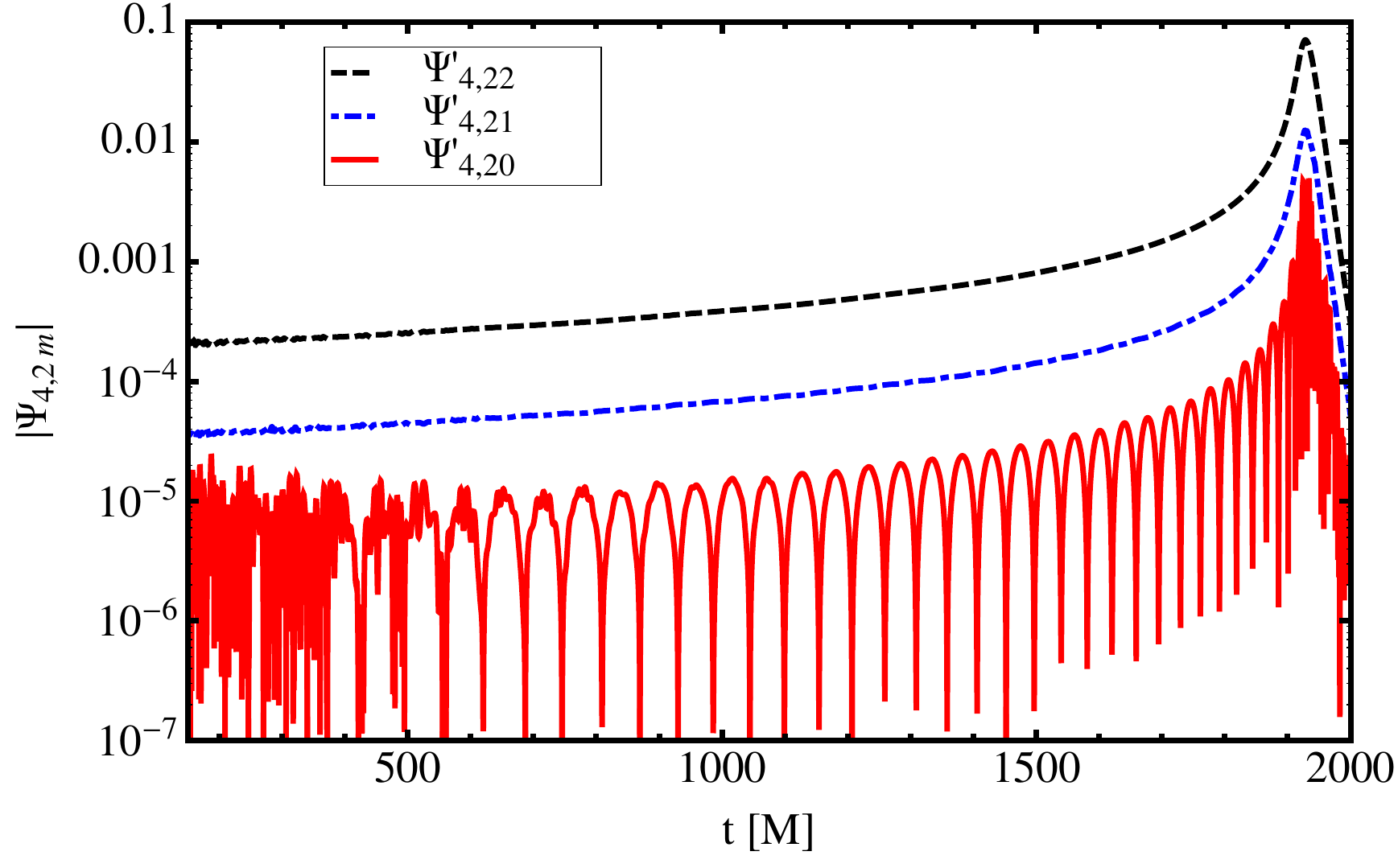}
\caption{
The left panel shows the amplitude of the $\tilde{\Psi}_{4,2m}$ modes for
the reference case. The $(\ell=2,m=1)$ mode is zero by symmetry,
and we see that the $(\ell=2,m=0)$ mode is much smaller than the 
dominant mode, and is essentially noise during most of the inspiral.
The right panel shows the corresponding amplitudes for the rotated 
case. We now see that both sub-dominant modes have become
significant. The amplitude of the $(\ell=2,m=0)$ mode is oscillatory
because it is a purely real function; see text for more details.
}
\label{fig:amps}
\end{figure*}

In the second simulation the orbital plane is rotated. The orbital plane is first rotated 
about the $y$-axis by $10^{\circ}$, and then around the $z$-axis by $25^{\circ}$. 
The motion of the punctures in both the reference and rotated cases 
is shown in Fig.~\ref{fig:twistmotion}. The modes of $\Psi_{4,\ell m}$ are now mixed, 
and the power in the $\Psi_{4,22}$ mode is distributed amongst the other 
$(\ell=2)$ modes. This can be seen in Fig.~\ref{fig:amps}. In the reference
case (denoted by $\tilde{\Psi}_{4,\ell m}$), the $(\ell=2,m=1)$ mode is zero 
by symmetry, and the $(\ell=2,m=0)$ 
mode is dominated by numerical noise. In the rotated case, however, both 
sub-dominant modes have become significant. Note that oscillations are
visible in the $(\ell=2,m=0)$ mode amplitude because it is a purely real function.

We now want to see if our maximization procedure can recover the waveform
from the reference simulation. In our procedure we search for a rotation of the 
system by the Euler
angles $(\beta,\gamma)$ such that the coefficient of the $(\ell=2,|m|=2)$ modes 
is maximized. 
If the method works, we will recover the Euler angles $(-10^{\circ},-205^{\circ})$,
which correspond to the rotation we have described\footnote{The Euler angle to 
reverse the twist is $-205^\circ$ due to the freedom in performing the rotation about the $y$-axis clockwise or counterclockwise.}.

\begin{figure}[t]
\centering
\includegraphics[width=80mm]{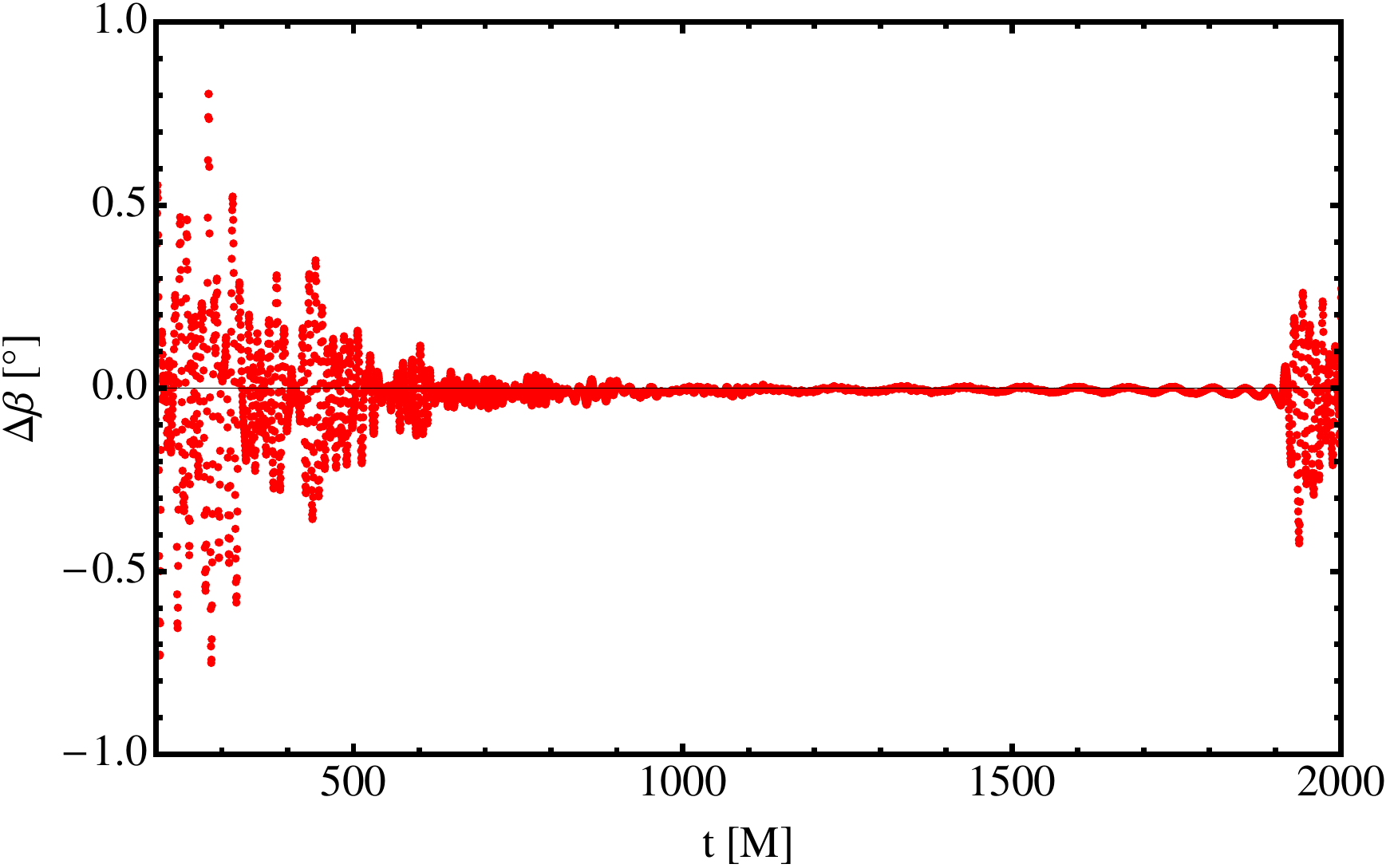}
\includegraphics[width=80mm]{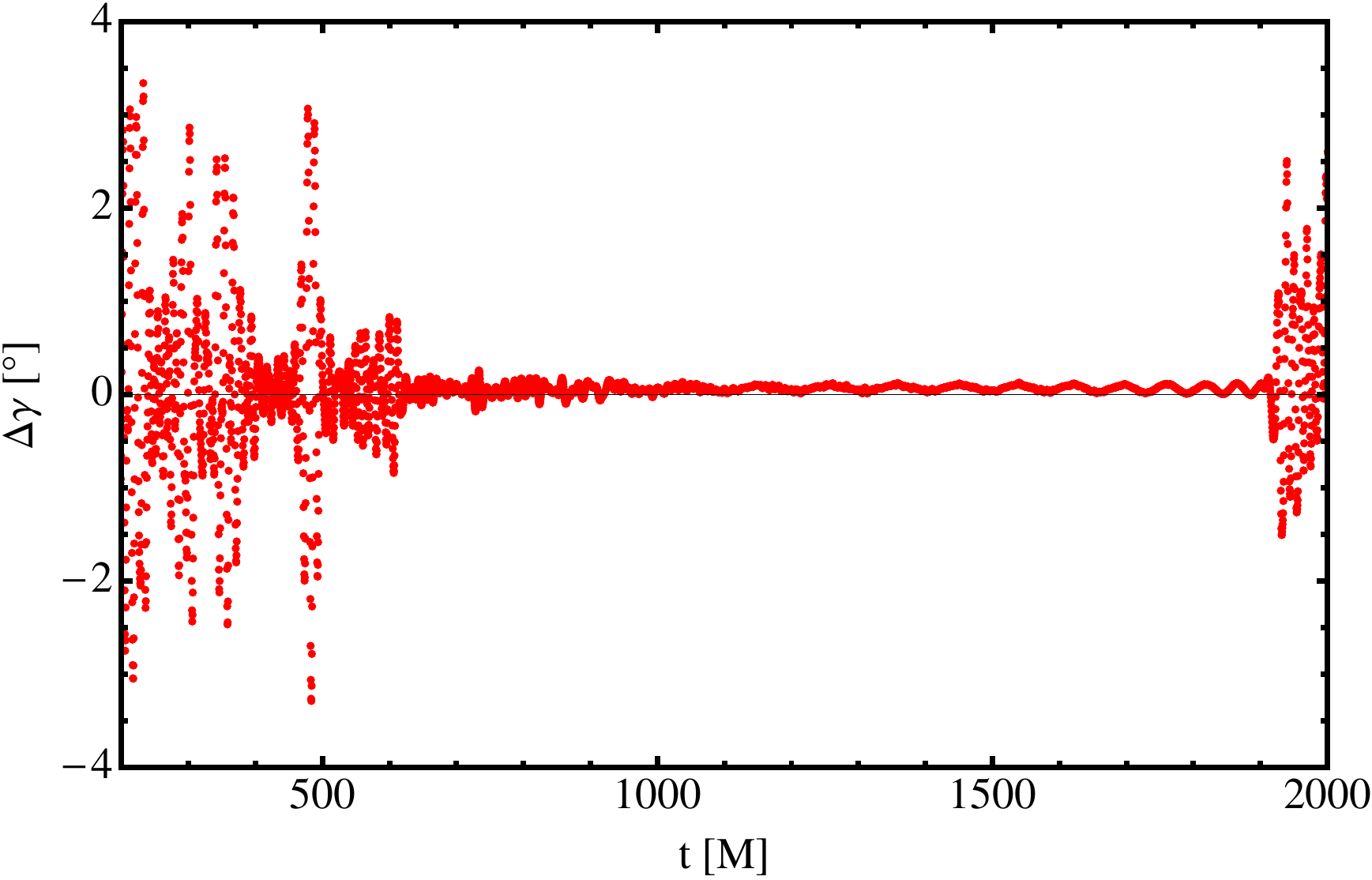}
\caption{
Error in the angles for the tilt ($\beta$) and twist ($\gamma$) of the orbital plane, 
as determined by the maximization procedure.
}
\label{fig:twistangles}
\end{figure}

Fig~\ref{fig:twistangles} shows the error in the determination of the Euler angles.
The maximization procedure was applied from $t=150M$, after the burst of junk 
radiation 
has passed, through to $t=2000M$, which is late in the ringdown phase. Up until 
about $t = 500M$ the waveform is rather noisy, and so the error in $\beta$ can be
as large as $1^{\circ}$, and in $\gamma$ the error is up to $4^{\circ}$. 
During most of the inspiral, however, when the wave signal
is clean, the error in $\beta$ is below $0.05^{\circ}$, and the error in $\gamma$
is below $0.2^{\circ}$, and even during ringdown (when the 
waveform amplitude is falling exponentially), the angles are determined to within
$\pm(0.5^{\circ},2.0^{\circ})$.

The magnitudes of the $\ell=2$ modes in the quadrupole-aligned waveform 
agree well with those in the reference case. The $(\ell=2,|m|=2)$ modes agreed
within numerical error in the raw data, 
and the $(\ell=2,|m|=1)$ modes, which should be zero by symmetry, 
were reduced by three orders of magnitude, to a level that would generally be
regarded as noise. During the inspiral, for example, $|\Psi_{4,21}|$ was reduced
from $\sim10^{-4}$ to $\sim10^{-7}$.

These results demonstrate that our method works, and give us an indication of the 
error bounds. We
expect that in general the errors will depend on the orientation angles of the 
binary, and will be worse when the angles are small. In these cases the 
sub-dominant modes will be smallest, and therefore will be resolved with 
less accuracy in the numerical code, and will then contribute more noise
to the waveform in the rotated frame. However, we will take the errors from 
this example as the basis for our error bounds in other applications of our
method.

\subsection{Precessing binary}

Having shown that the maximization procedure works for the equal-mass nonspinning 
test case, where the orientation of the orbital plane is known, we now apply the method
to a precessing binary. The configuration we have chosen has a mass ratio
of $q = M_2/M_1 = 3$, the larger black hole has a spin of $S_2/M_2^2 = 0.75$, and the 
spin initially lies in the orbital plane
, i.e., perpendicular to the Newtonian
orbital angular momentum. The small black hole is not spinning.

We expect this configuration to exhibit significant precession.  The leading
post-Newtonian contribution due to spin is the spin-orbit interaction, which can
be characterized by the Hamiltonian \cite{Kidder:1995zr} 
(see e.g. also \cite{Buonanno:2005xu})
 \begin{equation}\label{eq:HSO}
H_{\text{SO}} = 2 \frac{{\mathbf S_{\text{eff}}\cdot \mathbf L}}{r^3},
 \end{equation}
where $r$ is the coordinate separation of the black holes, and the effective 
spin ${\mathbf S}_{\mbox{\small eff}}$, which is defined as 
\begin{equation}
{\mathbf S}_{\mbox{\small eff}} = 
\left( 1 + \frac{3}{4}\frac{M_2}{M_1} {\mathbf S}_1 \right) +
\left( 1 + \frac{3}{4}\frac{M_1}{M_2} {\mathbf S}_2 \right),
 \end{equation}
where in our case one of the spins would be zero.
From the spin-orbit interaction one can derive a post-Newtonian
evolution equation for the black-hole spin~\cite{Kidder:1995zr},
\begin{equation}
\dot{\mathbf{S}} =
   - \frac{2}{r^3} {\mathbf S}_{\mbox{\small eff}} \times {\mathbf L}. 
\end{equation}
This indicates that the maximum amount of spin precession will be
achieved when the spin is perpendicular to the orbital angular momentum. 
If one of the black holes has a Kerr parameter $S_i/M_i^2$, then 
$S$ will be largest if the larger black hole is spinning. 
This is also convenient from a numerical point of view,
because the resolution requirements  increase both as the mass is decreased,
and spin is added; it is computationally cheaper to put the spin on the 
larger black hole. 

We also know from PN theory that 
$\dot{\mathbf{S}} = - \dot{\mathbf{L}}$ to leading order. 
If we increase the mass ratio, 
then the orbital angular momentum $L$ at a given separation will decrease, but the 
magnitude of the spin will stay the same. Therefore the relative change in $\mathbf{L}$ 
due to the precession of the spins will increase. This means that we will get greater
spin precession for higher mass ratios. We have chosen $q=3$ because this is
reasonably large compared to typical simulations we have performed in the past, but
low enough that we still expect to be able to achieve high accuracy. 

\begin{figure}[t]
\centering
\includegraphics[width=80mm]{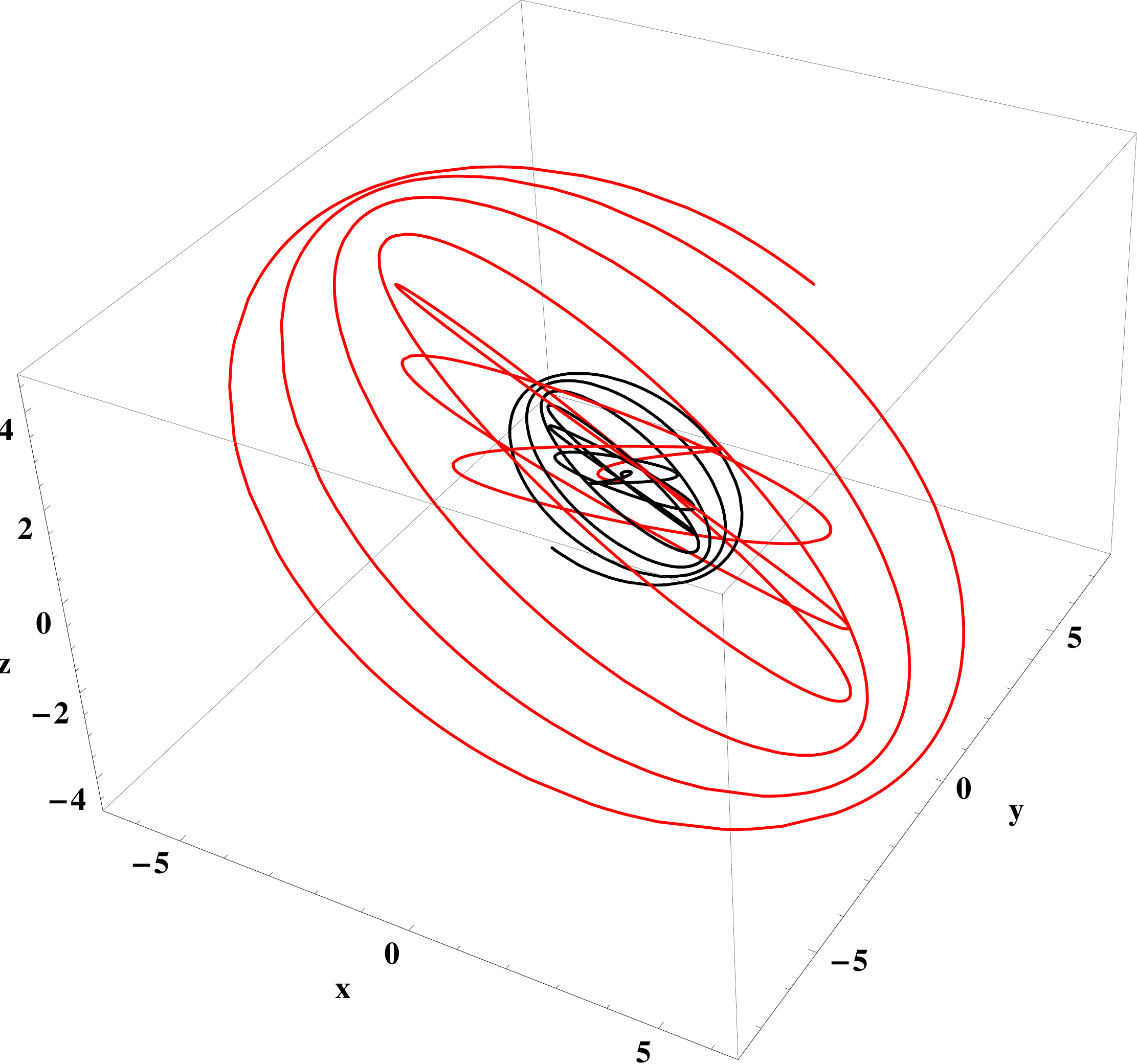}
\caption{
Motion of the black-hole punctures for the $q=3$ precession simulation. The motion of the 
small black hole is shown in red, and the large black hole is shown in black. 
The precession of the orbital plane is clearly visible through the inspiral. 
}
\label{fig:q3tracks}
\end{figure}

Fig.~\ref{fig:q3tracks} shows the orbital motion of the two punctures in the simulation.
The precession of the orbital plane is clearly visible in the figure. 

\begin{figure*}[t]
\centering
\includegraphics[width=80mm]{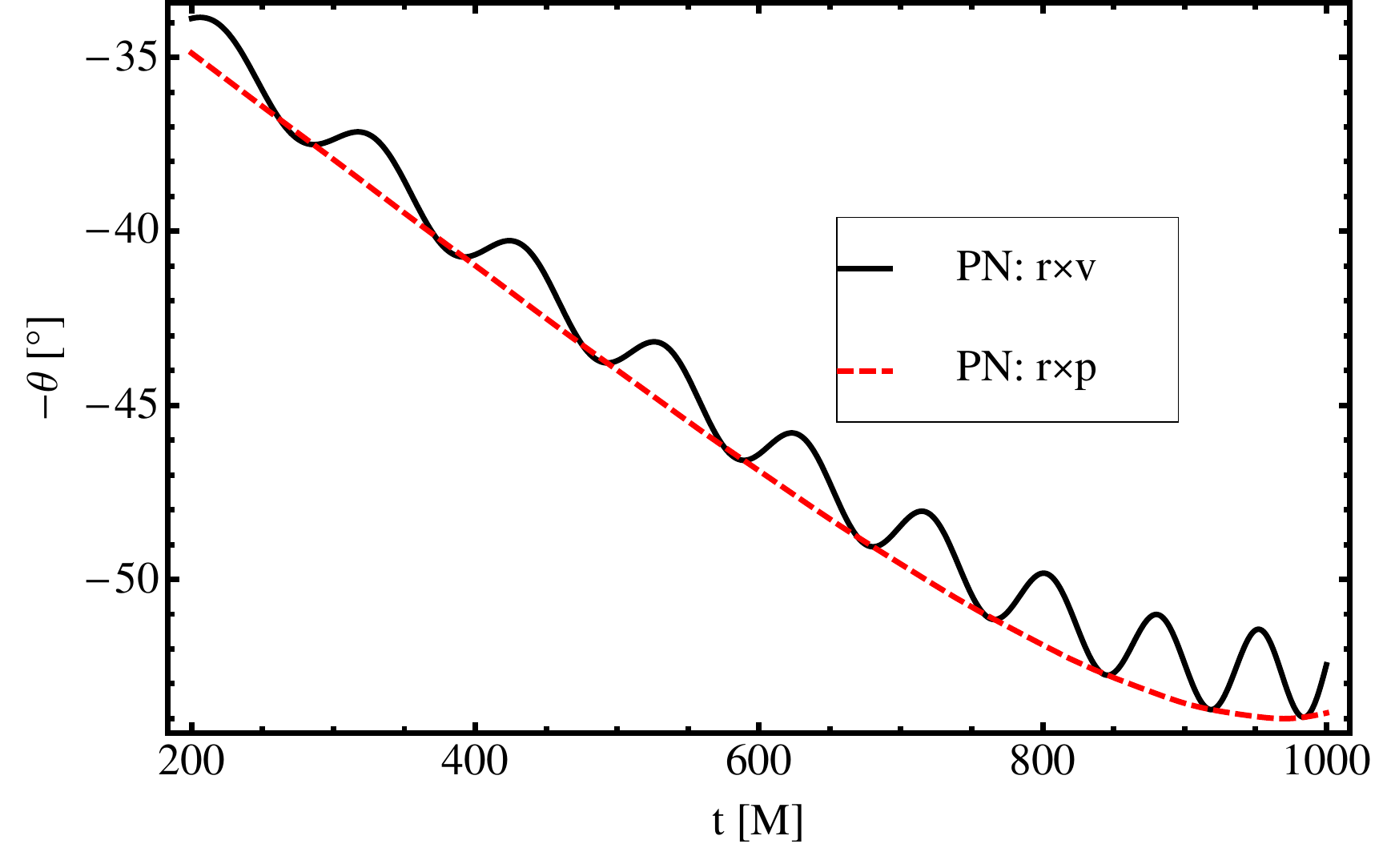}
\includegraphics[width=80mm]{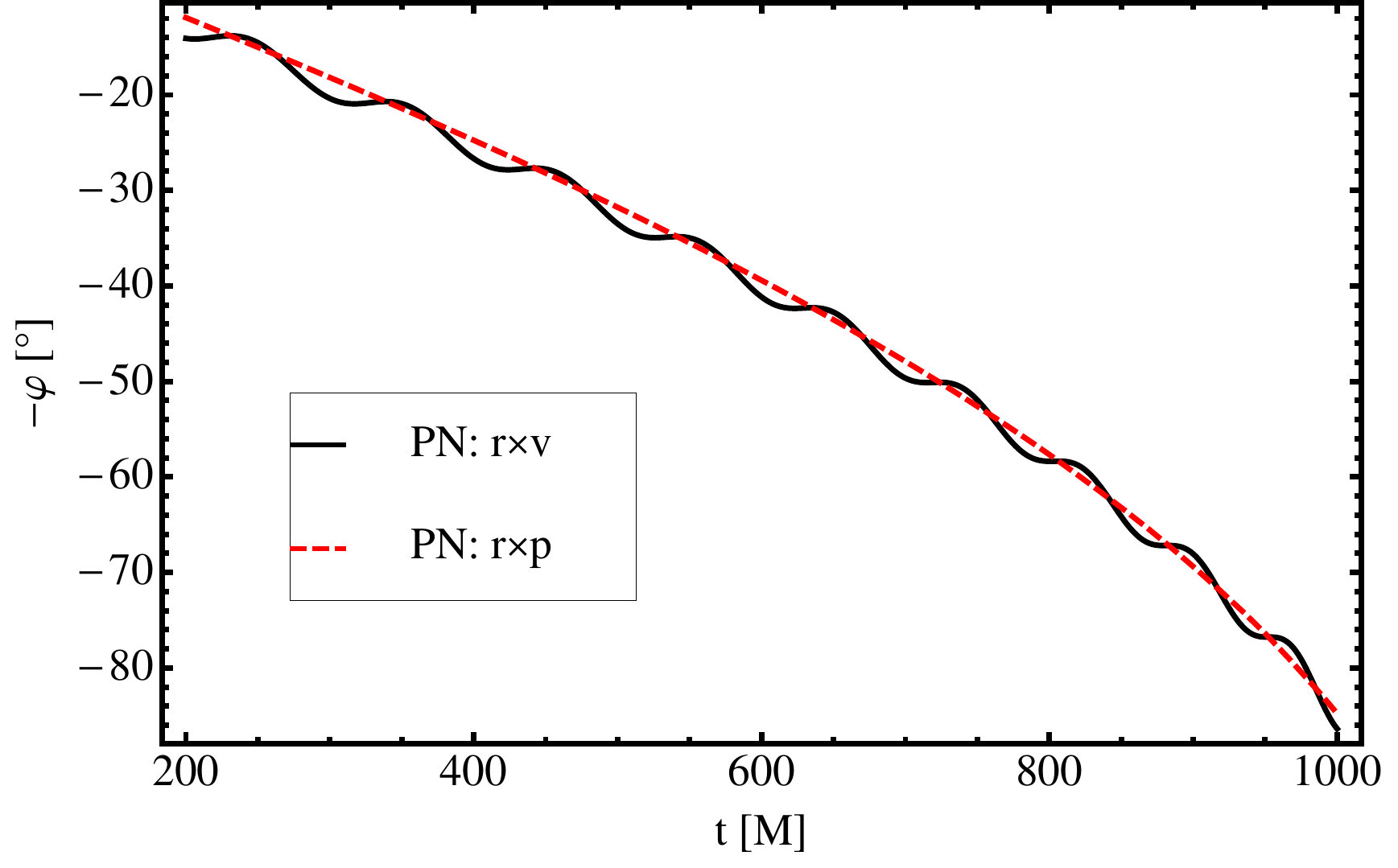}
\caption{
Comparison of the angles $\theta$ and $\varphi$ with respect to the 
$z$-axis for the directions of ${\mathbf r}\times {\mathbf v}$ (normal to the
orbital plane) and 
${\mathbf r}\times {\mathbf p}$ (orbital angular momentum) in 
a PN calculation. The comparison shows that the direction of 
${\mathbf r}\times {\mathbf v}$ exhibits extra oscillations.
}
\label{fig:AngleWiggles}
\end{figure*}

Considering the leading order spin-orbit interaction Eq.~(\ref{eq:HSO}) also exhibits another subtle
feature of spinning binaries. The time evolution of the momentum vector ${\mathbf p}$ is given
by the Hamiltonian evolution equation 
\begin{eqnarray}
\frac{d{\mathbf p}}{dt} = -\frac{\partial H}{\partial {\mathbf r}}.
\end{eqnarray}
If the Hamiltonian $H$ depends on the spins, then consequently the momentum also picks up
a contribution from the spins, and the velocity vector $\dot{\mathbf r}$ is in general {\em not}
parallel to the momentum ${\mathbf p}$. Consequently, the directions of the orbital frequency
vector $\boldsymbol \Omega$,
\begin{equation}
{\boldsymbol\Omega} = \frac{{\mathbf r}\times {\mathbf v}}{r^2}
\end{equation}
is in general {\em not} aligned with the angular momentum
${\mathbf L} = {\mathbf r}\times {\mathbf p}$.
For the spin-orbit interaction defined by the Hamiltonian in Eq. (\ref{eq:HSO}), this contribution to the angular momentum can be computed as
\cite{Kidder:1995zr}
\begin{eqnarray}\label{eq:LSO}
L_{\text{SO}} = \frac{\mu}{M} \left[ 
  \frac{M}{r}
      {\mathbf n} \times \left({\mathbf n} \times \left(3 {\mathbf S} + \frac{\delta m}{M} \Delta\right)\right) \right.  \nonumber \\
  \left. - \frac{1}{2}
      {\mathbf v} \times \left({\mathbf v} \times \left({\mathbf S} + \frac{\delta m}{M} \Delta\right)\right)\right],
\end{eqnarray}
where
\begin{equation}
L = L_{\text{NS}} + L_{\text{SO}},
\end{equation}
and $L_{\text{NS}}$ 
is the nonspinning contribution to the  angular momentum 
(which is parallel to the vector ${\mathbf r}\times {\mathbf v})$, ${\mathbf v} = \dot {\mathbf r}$, 
and ${\mathbf n}$ is the unit vector in the direction of ${\mathbf r}$.

Note that the effect of non-alignment  of ${\boldsymbol \Omega}$ and ${\mathbf L}$ is maximal 
when the spin ${\mathbf S}$ is in the orbital plane. This is indeed the case for our initial conditions,
and also during the  evolution the spin component out of the orbital plane is significantly smaller
than  the components in the  orbital plane.  Note also that since the spin typically varies on a timescale
larger than the orbital time scale, Eq.~(\ref{eq:LSO}) will lead to oscillations in the angle between
${\boldsymbol \Omega}$ and ${\mathbf L}$  with roughly the orbital period.

Such oscillations are not present in the direction of $\mathbf{L}$,
as illustrated in Fig. \ref{fig:AngleWiggles}. We will see
later that the quadrupole-aligned frame moves consistently with $\mathbf{L}$ 
(i.e,. as a smooth function), suggesting that our maximization procedure tracks the
direction of the orbital angular momentum.

\begin{figure*}[t]
\centering
\includegraphics[width=80mm]{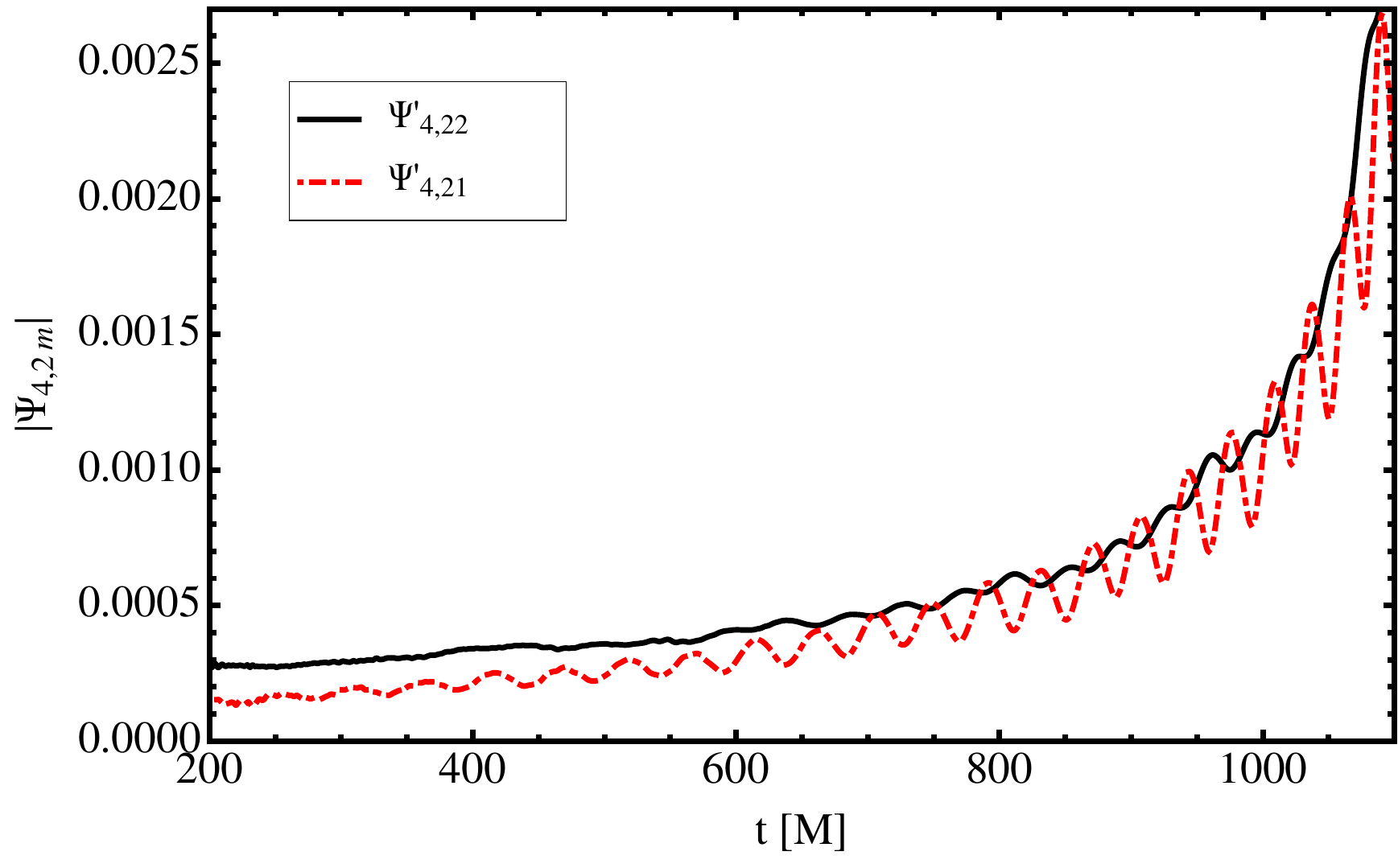}
\includegraphics[width=80mm]{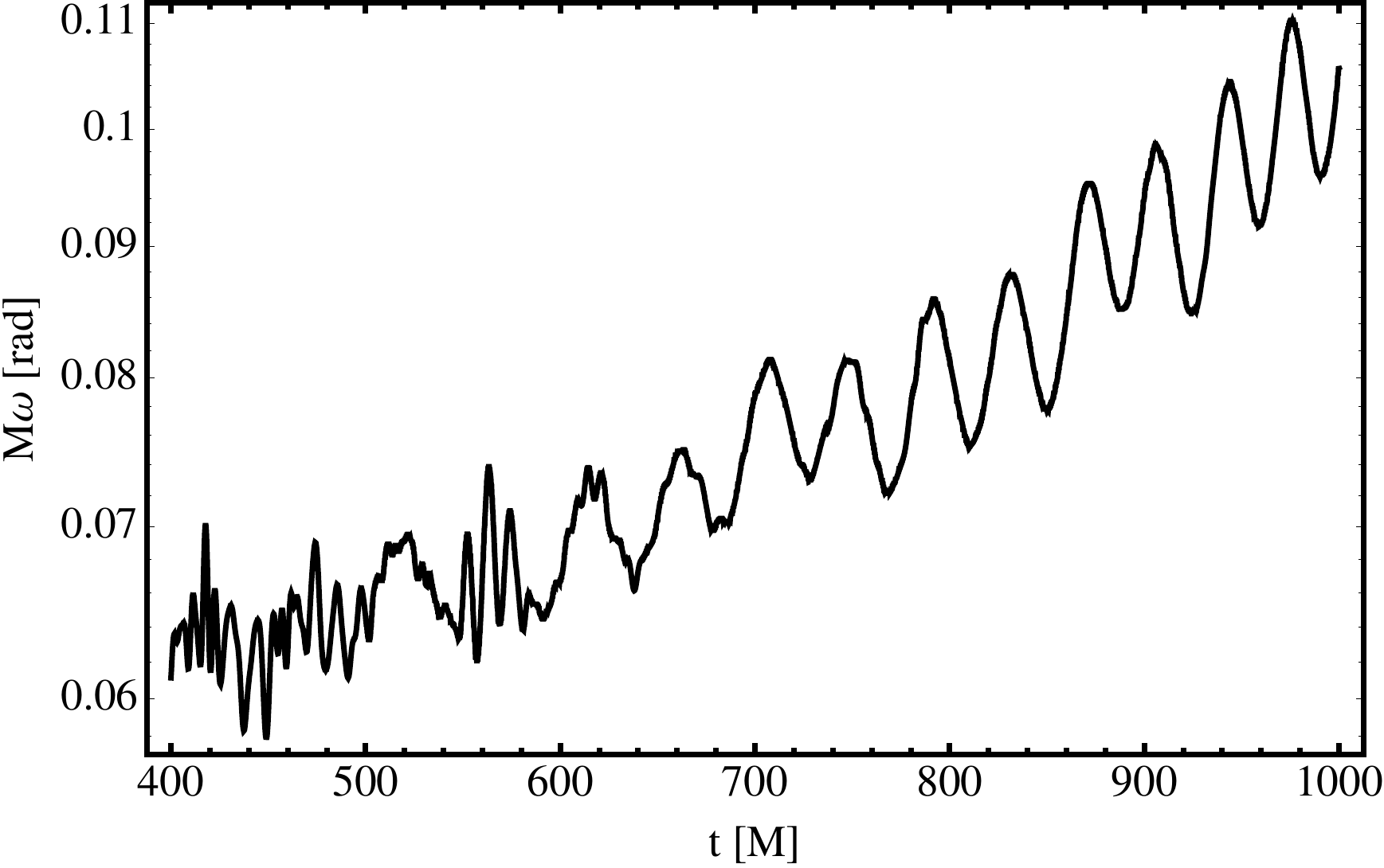}
\caption{
Amplitude of raw numerical data for inspiral (left), for the ``dominant''
mode $\Psi'_{4,22}$ and the ``sub-dominant'' mode $\Psi'_{4,21}$. 
The right panel shows the frequency of the $(\ell=2,m=2)$ mode, 
which exhibits significant oscillations. (The data are also noisy at 
early times, but this is typical for such data.) 
}
\label{fig:q3InspiralRaw}
\end{figure*}

The left panel of Fig.~\ref{fig:q3InspiralRaw} shows the amplitude of the $(\ell=2,m=2)$ and 
$(\ell=2,m=1)$ modes during the inspiral. We clearly see that the ``sub-dominant'' $(2,1)$
mode is of comparable magnitude to the $(2,2)$ mode, and shows significant
modulation. (It is also instructive to compare with the results in~\cite{Campanelli:2008nk}, 
where a precessing binary is also considered from a fixed frame of reference, and all of 
the $\ell=2$ modes are of significant amplitude.)
The right panel of Fig.~\ref{fig:q3InspiralRaw} shows the frequency of the 
$(2,2)$ mode, $\omega_{22} = \dot\varphi_{22}$, over the same time interval. The
frequency clearly exhibits large oscillations. 
Based on the discussion around Eq.~(\ref{eqn:freqreln}) we expect 
oscillations in $\omega_{22}$ of purely physical origin, but we also assume
that the physical oscillations will be exaggerated and their frequency modified
in the fixed frame of an inertial observer.

We now apply the maximization procedure to the waveform signal from 
$t=150M$, when the junk radiation has passed, through merger
and ringdown (up to $t=1250M$). At each time step the system is rotated
such that the $(\ell=2,|m|=2)$ mode amplitudes are maximized.

Finally, we address the question of whether the GW signal is emitted normal to 
the orbital plane, or parallel to the orbital angular momentum. Although we 
cannot unambiguously define the direction of orbital angular momentum, we
can certainly determine whether the GW signal is emitted normal to the
orbital plane.

\begin{figure*}[t]
\centering
\includegraphics[width=80mm]{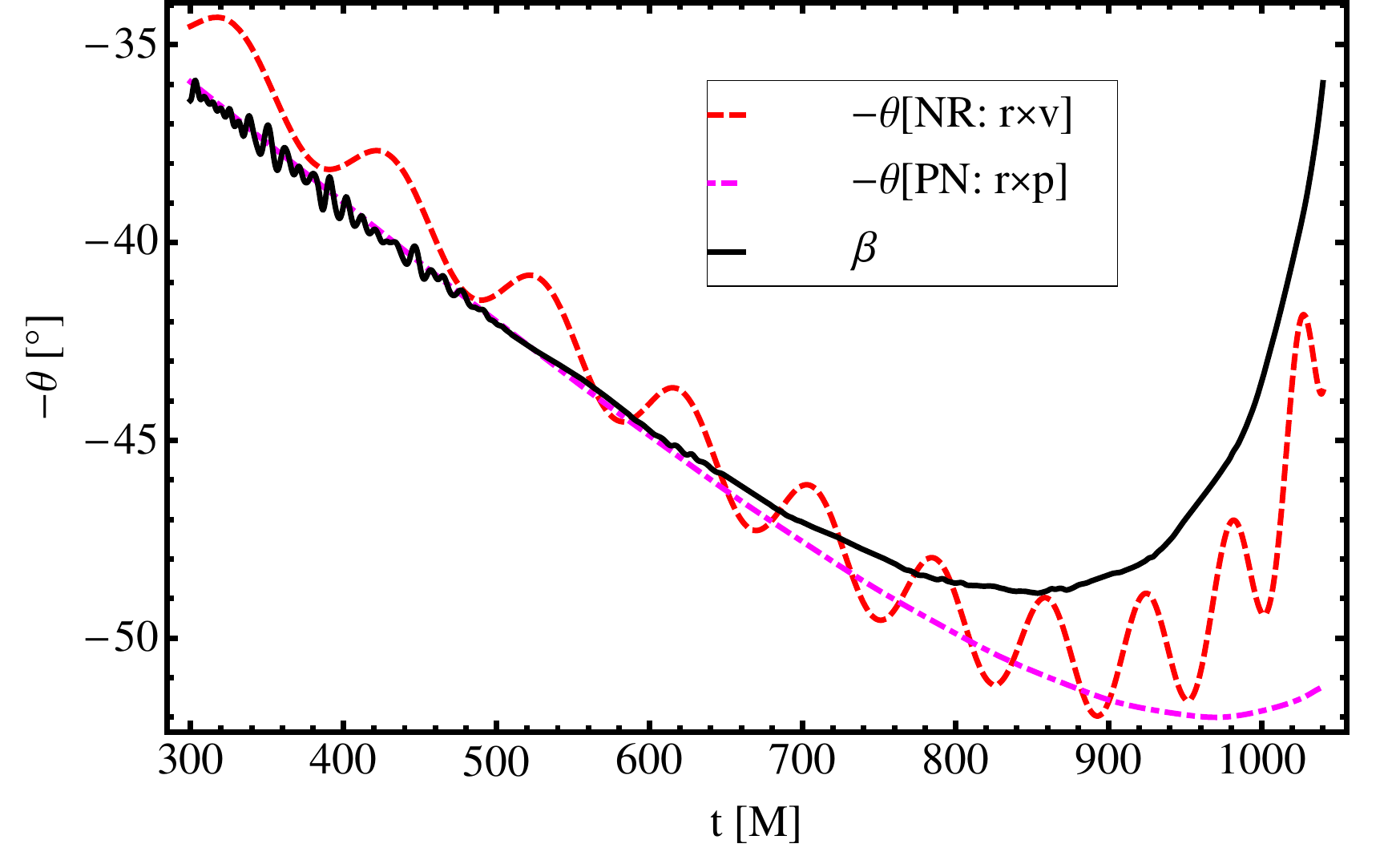}
\includegraphics[width=80mm]{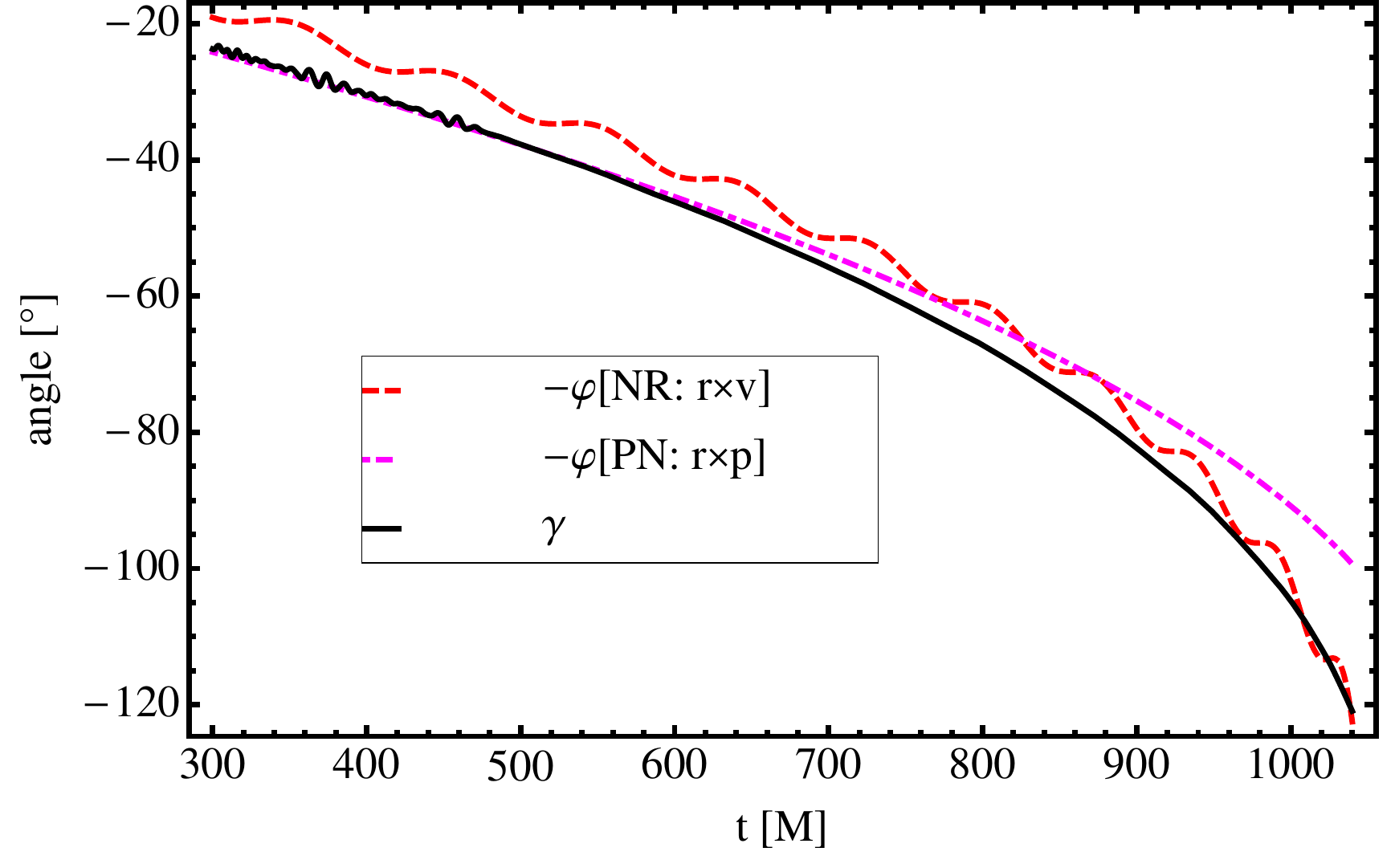}
\caption{
The Euler angles $(\beta,\gamma)$ found when the maximization procedure
was applied to the $q=3$ precessing-binary waveform. For comparison we show the
corresponding angles $(-\theta,-\varphi)$ of the normal to the orbital plane as
computed from the NR simulation, and for the angular momentum ${\mathbf L}$ from
a PN simulation (as in Figs.~\ref{fig:AngleWiggles}). We apply
an appropriate time shift to $(\beta,\gamma)$ 
to approximately compensate for
the time lag at the wave extraction sphere, and add $2 \degree$ to the curves
for $\theta$ as described in the text. We clearly see
that the orbital-plane angles show additional oscillations that are not 
present in the (2,2)-maximization angles. 
}
\label{fig:q3angles}
\end{figure*}

Fig.~\ref{fig:q3angles} shows the Euler angles $(\beta,\gamma)$ that were 
found in the maximization procedure, time shifted by $100M$ to approximately compensate for
the time lag to the extraction spheres. It also shows the angles $(\theta,\varphi)$ 
of the direction orthogonal to the orbital plane as computed from the NR simulation,
and  for the orbital angular momentum ${\mathbf L}$ as computed from
a PN simulation (as in Figs.~\ref{fig:AngleWiggles}).
A constant offset of $2 \degree$ has been
added to $\theta$ to align the $\theta$ angles and $\beta$. 
A potential explanation for this small offset are coordinate gauge ambiguities.
If the GW signal were emitted normal to the orbital 
plane, we would expect to be able to align $\beta$ with $-\theta$ from the numerical
relativity simulation, and likewise for
 $\gamma$ and $-\varphi$. However, it is clear from Fig.~\ref{fig:q3angles} 
that the orbital-plane angles contain extra oscillations. Based on the 
illustration in Fig.~\ref{fig:AngleWiggles}, 
this suggests that the GW
signal is emitted in the direction of the orbital angular momentum. In particular,
show in Fig.~\ref{fig:q3angles} the direction of the orbital angular momentum as
predicted in PN theory shows
good agreement with the angles that define the quadrupole-aligned frame.

\begin{figure*}[t]
\centering
\includegraphics[width=80mm]{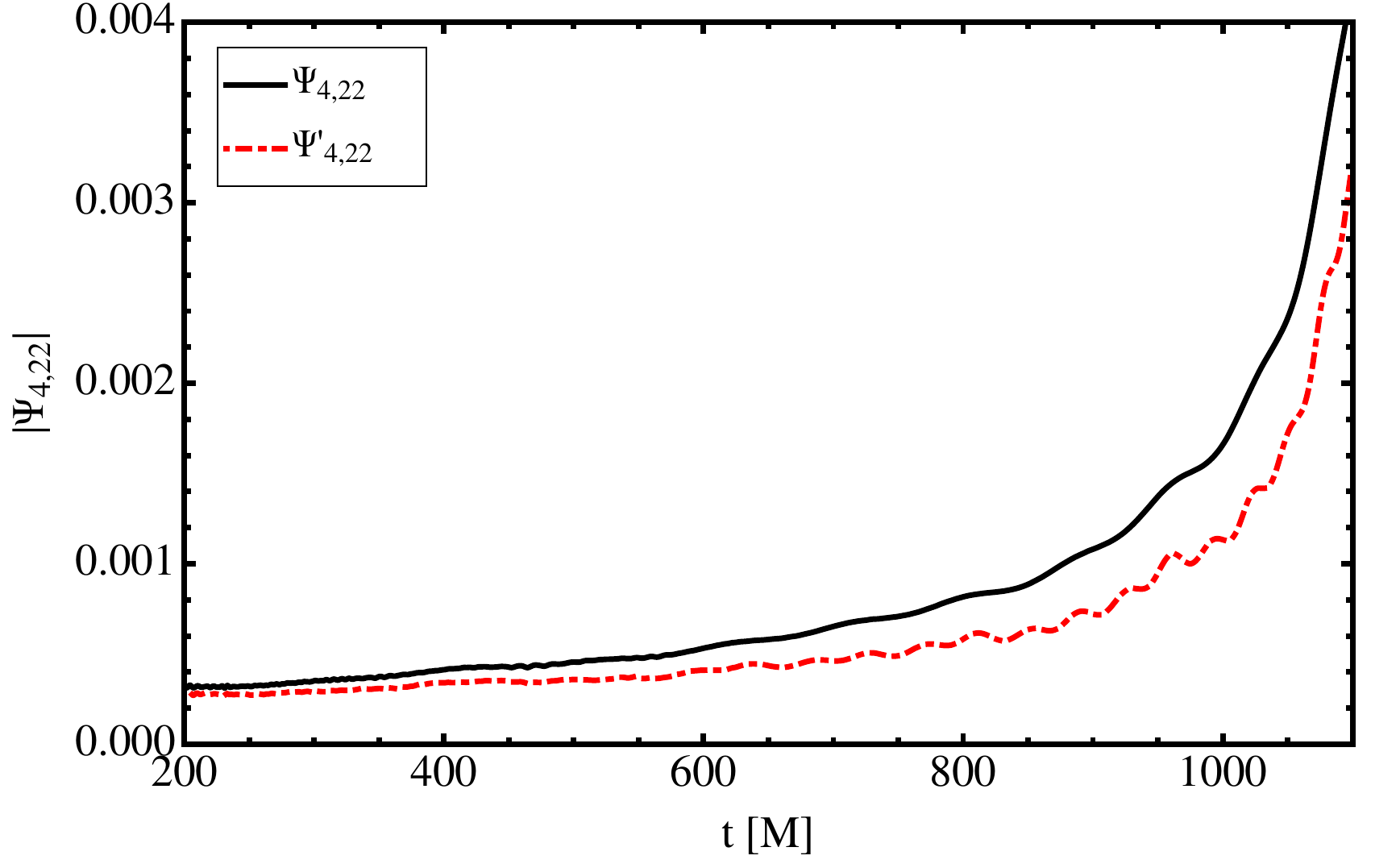}
\includegraphics[width=80mm]{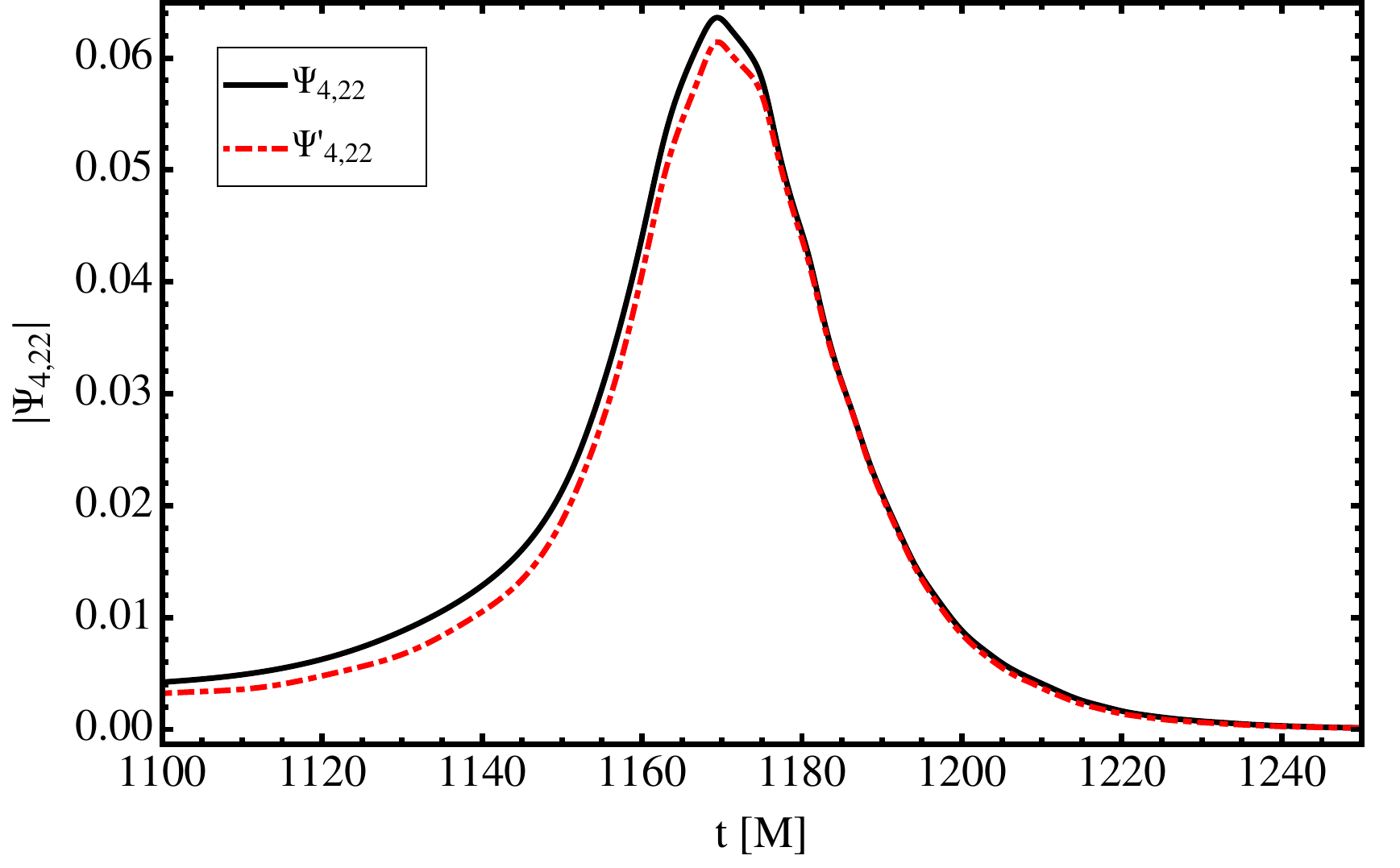}
\caption{~
Amplitude of the $(\ell=2, m=2)$ mode, before ($\Psi'_{4,22}$)
and after ($\Psi_{4,22}$) the maximization procedure.
}
\label{fig:ReconstructedAmp}
\end{figure*}

Fig.~\ref{fig:ReconstructedAmp} shows the amplitude of the original 
$\Psi'_{4,22}$ 
and the quadrupole-aligned signal that results from the maximization 
procedure, $\Psi_{4,22}$. We see that the maximization procedure has
indeed increased the amplitude at all times, and also seems to have removed
some oscillations.

\begin{figure}[t]
\centering
\includegraphics[width=80mm]{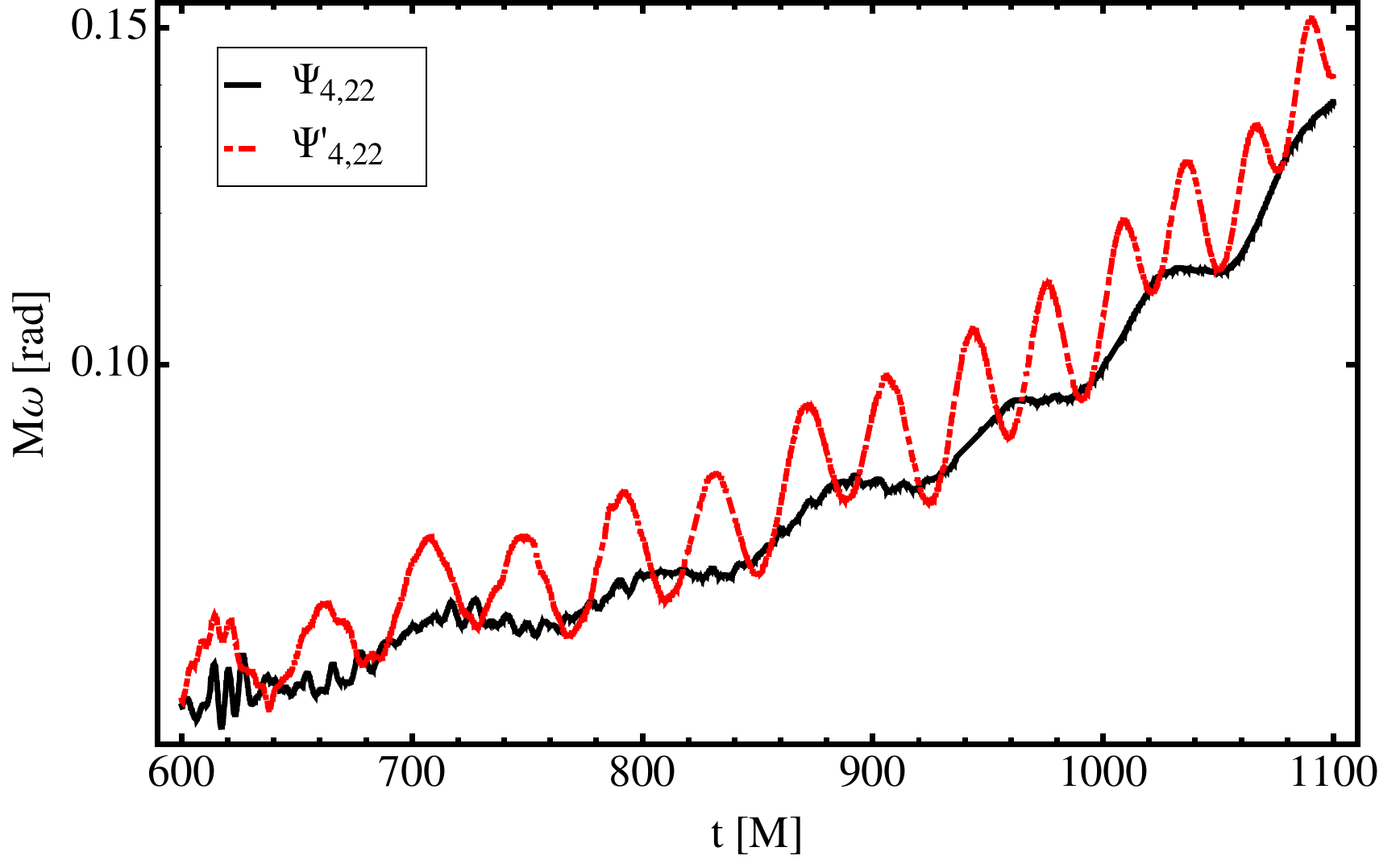}
\caption{
Frequency of the $(\ell=2,m=2)$ mode before ($\Psi'_{4,22}$) and 
after ($\Psi_{4,22}$) the maximization procedure. We see that the 
high-frequency oscillations have been removed. 
The remaining oscillations are of a lower frequency and much lower 
amplitude; see text and Fig.~\ref{fig:frequency_comparison_with_Arun}.
}
\label{fig:frequency}
\end{figure}

\begin{figure}[t]
\centering
\includegraphics[width=80mm]{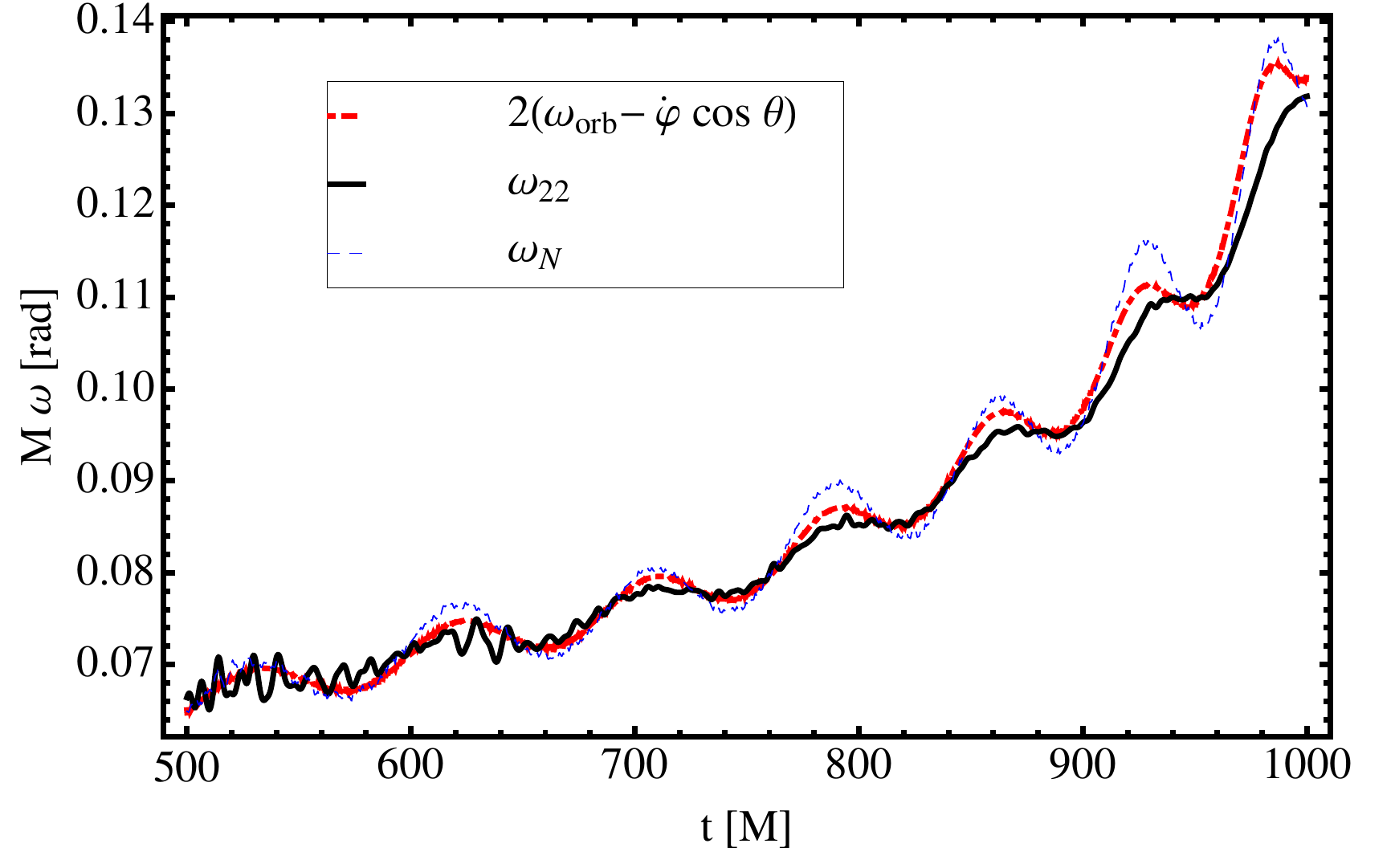}
\caption{
Frequency of the $(\ell=2,m=2)$ mode 
after ($\Psi_{4,22}$) the maximization procedure, compared with
the orbital frequency with a precession term added according
to Eq.~(\ref{eqn:freqreln}). We also show the frequency that results
from rotating the system according to the direction of the Newtonian 
orbital angular momentum, $\omega_N$, i.e., the normal to the orbital
plane. 
}
\label{fig:frequency_comparison_with_Arun}
\end{figure}

The frequency of the ($\ell=2,m=2)$ mode before and after the 
maximization procedure is shown in Fig.~\ref{fig:frequency}. This figure
illustrates the key result of this work: the high-frequency oscillations in
the wave frequency have been removed by the maximization procedure,
and we are left with a far simpler functional form. We note, however, that the 
oscillations in the frequency have not been completely removed. 
This is to be expected from Eq.~(\ref{eqn:freqreln}).
In the absence of precession, during the inspiral the gravitational wave frequency
of a spherical harmonic mode $(\ell,m)$ is with a high degree of accuracy
proportional to the orbital frequency,  $\omega_{\ell m} = m \omega_{\text{orb}}$.
In the  presence of precession, this is however replaced by Eq.~(\ref{eqn:freqreln}),
which adds an extra term depending on the precessing motion of the orbital plane.
In Fig.~\ref{fig:frequency_comparison_with_Arun} we compare the frequency of the $(\ell=2,m=2)$ mode
after the maximization procedure with
the orbital frequency with the precession term added according
to Eq.~(\ref{eqn:freqreln}), and we find reasonable agreement. We also show the
frequency $\omega_N$ that results from rotating the system according to the direction 
perpendicular to the orbital plane, which is also the direction of the naive Newtonian 
orbital angular momentum. 
It is clear from Fig.~\ref{fig:frequency_comparison_with_Arun} that the oscillations due to the 
orbital-plane rotations are much larger, and this further suggests that the quadrupole-aligned
frame is optimal.

It is clear that the maximization procedure produces $(\ell=2,|m|=2)$
modes that are of a simpler form than in the original data. 
However, this is not a guarantee that we have correctly
tracked the direction of the GW emission; we have not necessarily put the
waveform into a physically meaningful frame of reference. One test of 
our method is to calculate the effect on the sub-dominant modes. We
expect that in the quadrupole-aligned frame the amplitude of the GW 
signal will agree to a good approximation with that from a $q=3$
nonspinning binary. (The spin effect on the rate of inspiral is dominated 
by $\mathbf{S} \cdot \mathbf{L}$, and this is close to zero throughout
our simulation, so we expect the inspiral to be similar to that for a 
nonspinning binary with the same mass ratio.)

\begin{figure*}[t]
\centering
\includegraphics[width=80mm]{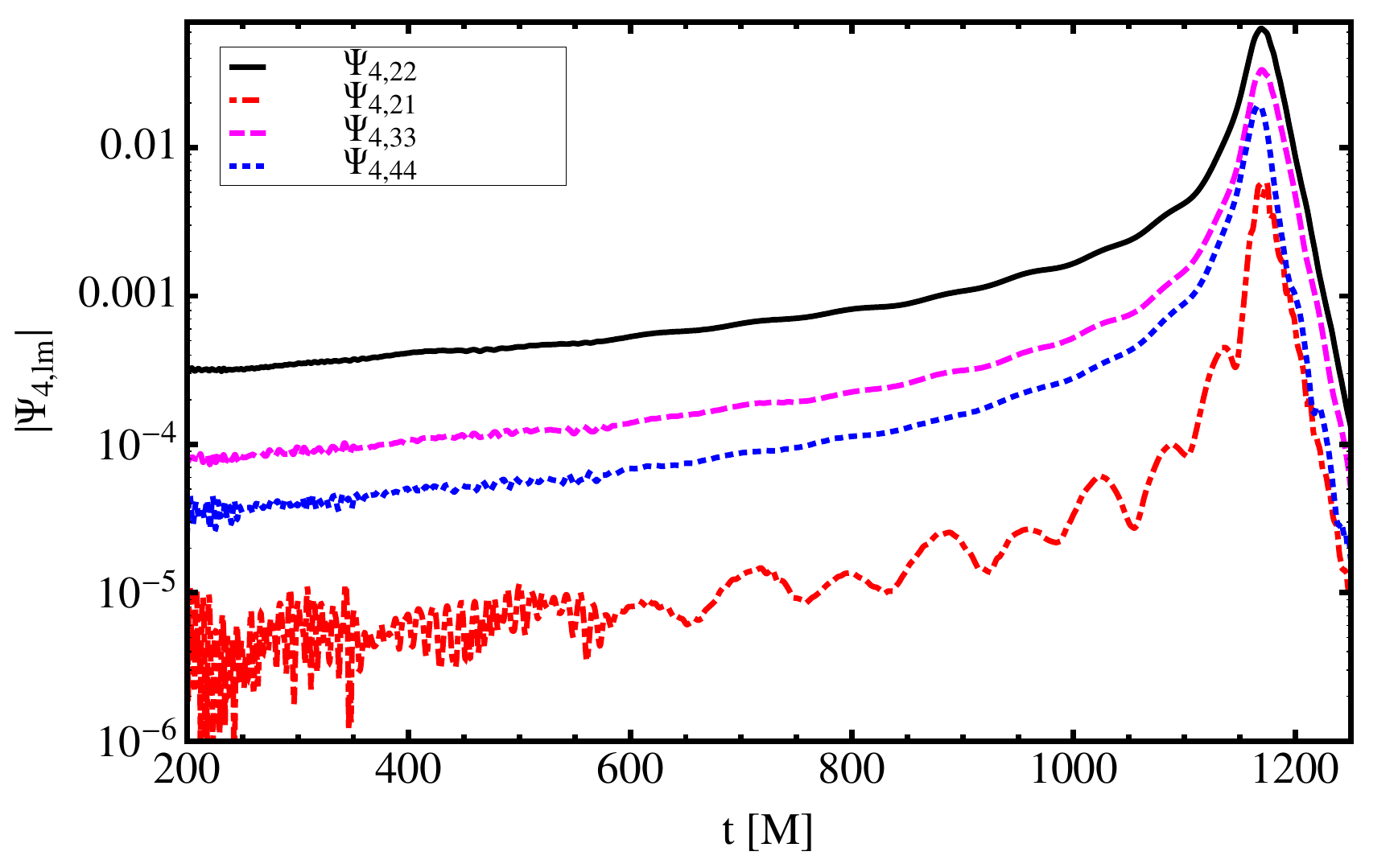}
\includegraphics[width=78mm]{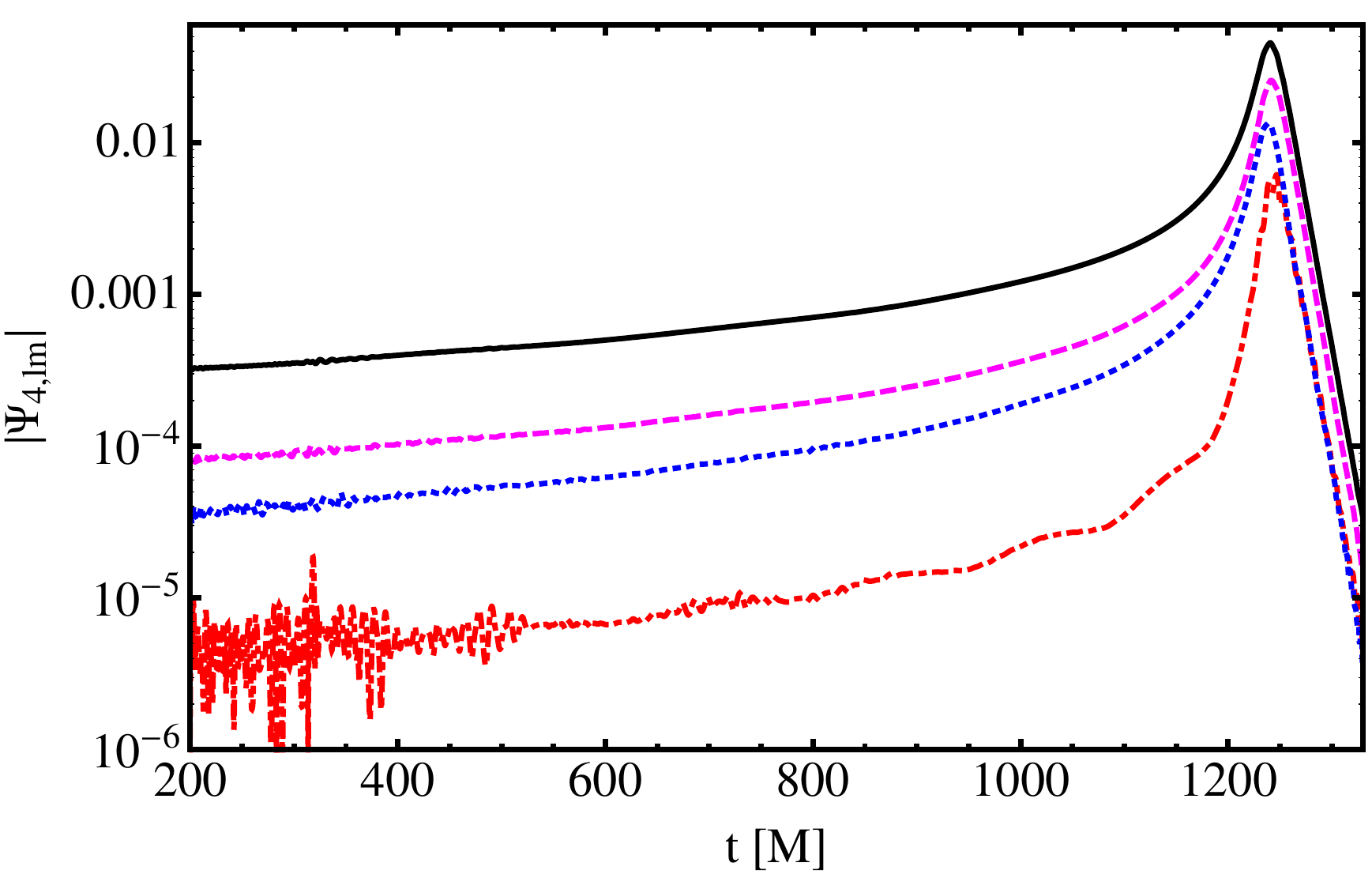}
\caption{
Left: selected modes of the precessing-binary waveform, after being transformed
into the non-precessing frame, i.e., after the system has been rotated by the 
angles that were found from the (2,2)-maximization procedure. The right-hand
plot shows the same modes for a nonspinning (and therefore non-precessing)
$q=3$ waveform. The agreement is remarkable. Note in particular the qualitative
agreement of the $(\ell=2,m=1)$ mode, which is of comparable magnitude to the
$(\ell=2,m=2)$ mode in the raw data (see Fig.~\ref{fig:q3InspiralRaw}). 
}
\label{fig:newmodes}
\end{figure*}

Fig.~\ref{fig:newmodes} shows a selection of modes for the quadrupole-aligned
waveform. The left frame shows the transformed modes for
the precessing binary, and the right frame shows the same modes for
the nonspinning $q=3$ waveform presented in~\cite{Hannam:2010ec}. 
Two things are remarkable about this figure. The first is that the amplitudes
of the modes show extremely good agreement. The other is that 
we have found that the magnitude of the $(\ell=2,m=1)$ mode is extremely 
sensitive to the angle by which the system is rotated. If, for example, we were
to modify $\beta$ or $\gamma$ by a fraction of a degree, $\Psi_{4,21}$ could
change by orders of magnitude. With this fact borne in mind, the oscillations in 
$|\Psi_{4,21}|$ are not very large at all. This figure suggests that we have
located an optimal frame from which to study the GW signal.

\section{Discussion}
\label{sec:conclusion}

We have presented a simple method to track the precession of a binary system,
using only information from the GW signal. Our procedure is to rotate the system
such that the magnitude of the $(\ell=2,|m~|=2)$ modes is maximized, based on the 
physical assumption that this will be the direction of dominant GW emission. 
We refer to this as the ``quadrupole-aligned'' waveform.
Based on evidence from PN theory, we show that this direction seems to 
correspond to that of the orbital angular momentum, which is in general
\emph{not} perpendicular to the orbital plane. We also show that our method 
produces higher-mode amplitudes consistent with what we know from 
non-spinning, non-precessing binaries. 

The result of our procedure is that the waveform is represented in a more
simple form than the one produced directly from the numerical code. This
is particularly true of the subdominant modes; compare Figs.~\ref{fig:q3InspiralRaw}
and \ref{fig:newmodes}. We expect
that this will simplify the task of producing analytic inspiral-merger-ringdown
models, which is one of the main motivations for our work. 
This method also provides a \emph{normal form} for 
the waveform, which will facilitate future comparisons between numerical and 
analytic results.

One could propose alternative procedures to track the precession of the system,
and we will now discuss some of them, and their difficulties.

Only the total angular momentum of the spacetime is unambiguously defined in 
General Relativity. The form of Bowen-York puncture initial data are such that 
we can analytically calculate the angular momentum 
(\cite{OMurchadha1974,York:1974hp,York79,Brugmann:2008zz})
of the initial slice from the initial-data parameters; it is simply given by 
$\mathbf{L} = \mathbf{r}_1 \times \mathbf{p}_1 +  \mathbf{r}_2 \times \mathbf{p}_2$,
where $\mathbf{r}_i$ are the coordinate locations of the punctures, and 
$\mathbf{p}_i$ are the momenta that are input into the Bowen-York extrinsic 
curvature. We can calculate the angular momentum radiated through the 
spheres on which we measure the GW signal, and so we can determine the
total angular momentum of the system as a function of time. However, we
want the \emph{orbital} angular momentum, $\mathbf{L = J - S}$. To 
calculate this we need to know the black-hole spins as a function
of time (which can be estimated with reasonable accuracy from the black
holes' apparent horizons~\cite{Campanelli:2006fy}), but these quantities are
calculated at the black holes, not at the GW extraction sphere, and cannot
easily be translated. 

One could attempt to instead calculate the orbital angular momentum entirely 
at the sources, but this also presents difficulties. 
The proper distance between the black-hole horizons and their
momenta could be calculated by some quasi-local procedure (for example 
\cite{Krishnan:2007pu}), and hence the orbital angular momentum.
But it will be difficult to assess the gauge errors in any such method. 
Alternatively, one could calculate the angular momentum using the puncture locations
and PN theory, but this will only be an approximation to the true general relativistic 
angular momentum. One direction we can easily determine is the normal to the orbital
plane of the binary, but we have seen in Sec.~\ref{sec:results}, that this is not the
direction in which the dominant GW signal is emitted, and nor does it define a reference
frame from which the GW signal appears simpler than what can be achieved by the 
maximization procedure that we have used.

Nonetheless, a number of issues remain to be resolved in our procedure. 
In particular, our method does not seem to 
accurately track the quadrupole-aligned direction through 
merger and ringdown. If it were able to do this, it would provide an alternative procedure
to determine the direction of the spin of the final black hole. We find that the angles
from our maximization procedure continue to vary through merger and ringdown, 
and do not settle at constant values, which is what they would do if they had
determined the final spin direction. This problem may be due to the accuracy of
the numerical data, or to more subtle effects, for example the motion of the 
center-of-mass of the system due to gravitational recoil. We will investigate
this issue further in future work. 

While working on this project we have learned that an independent effort 
to identify precession effects via a similar algorithm will be presented 
by Seiler et al. in a forthcoming publication \cite{Seiler:precess2}.

\section*{Acknowledgments}

We thank Jennifer Seiler for letting us know about her ongoing work on
a similar algorithm,
and
B.S. Sathyaprakash, Stephen Fairhurst,  Michael P\"urrer and Denis
Pollney for discussions.
P. Schmidt was partially supported by FWF grant P22498.
M. Hannam was supported by FWF grant M1178
and Science and Technology Facilities Council grants ST/H008438/1
and ST/I001085/1.
S. Husa was supported by
grant FPA-2007-60220 from the Spanish Ministry of Science and
the Spanish MICINN’s Consolider-Ingenio 2010 Programme under grant 
MultiDark CSD2009-00064, and thanks Cardiff University for hospitality.
P. Ajith was supported in part by NSF grants PHY-0653653 and 
PHY-0601459, and the David and Barbara Groce Fund at Caltech.
{\tt BAM} simulations were carried out at LRZ Munich, ICHEC Dublin,
the Vienna Scientific Cluster (VSC), and at MareNostrum at Barcelona 
Supercomputing Center -- Centro Nacional de Supercomputaci\'on 
(Spanish National Supercomputing Center).

\appendix

\section{Transformation of $\Psi_{4,lm}$ under rotations}
\label{sec:appendix}

We aim to derive the transformation of the Weyl scalar $\Psi_4$ under a rotation $\mathbf{R}\in SO(3)$. It can be shown that the Weyl scalar is  a field of spin-weight $s=-2$ and hence it can be expanded as
\begin{equation}
\label{eq:A1}
\Psi_4=\sum_{l,m}\Psi_{4,lm}Y^{-2}_{lm},
\end{equation}
where $Y^{-2}_{lm}$ denote the spherical harmonics of spin-weight $s=-2$ \cite{Newman:1966}. For $s=0$ we obtain the regular spherical harmonics $Y_{lm}$, which are
the eigenfunctions of the angle-dependent part of the Laplace operator.

The transformation of the spin-weighted spherical harmonics is a simple composition of the transformation of the spin-basis-dependent part and of $Y_{lm}$. It is convenient to introduce standard polar coordinates 
$(r,\theta,\varphi)$ and to define $Y_{lm}$ with respect to the polar angles $(\theta,\varphi)$.
The spherical harmonics then have the form
\begin{equation}
\label{eq:A2}
Y_{lm}(\theta,\varphi)=\phi(\varphi)\Theta(\theta).
\end{equation}

We will consider rotations $\mathbf{R}$, which transform angles 
$\Omega=(\theta,\varphi)$ to the new coordinates $\Omega'=(\theta',\varphi')$.
The spin-weight-zero spherical harmonics $Y_{lm}$ then transform according to
 $Y_{lm}(\theta,\varphi)\mapsto Y_{lm}(\theta',\varphi')$ by applying the operator $\mathbf{P}_R$, where $R$ is a rotation about the z-axis by the angle $\gamma$ such that $\varphi\mapsto \varphi'=\varphi+\gamma$ and $\theta=\theta'$, is given by 
\begin{equation}
\label{eq:A4}
Y_{lm}(\theta',\varphi')\equiv\mathbf{P}_R Y_{lm}(\theta,\varphi)=e^{im\gamma}Y_{lm}(\theta,\varphi).
\end{equation}
Now, let $\mathbf{R}(\gamma\beta\alpha)$ denote an arbitrary rotation by the Euler angles $\gamma,\beta,\alpha$.
Using the z-y-z convention, the spherical harmonics then obey the following transformation law \cite{Wigner1959, Goldberg:1967}:
\begin{equation}
\label{eq:A5 }
Y_{lm}(\theta',\varphi')=\sum_{m'=-l}^l e^{im'\gamma}d^l_{m'm}(\beta)e^{im\alpha}Y_{lm}(\theta,\varphi),
\end{equation}
where the $d^l_{m'm}$ denote the Wigner $d$-matrices which are given by
\begin{align}
\label{eq:Wigner}
d^l_{m'm}=&\sqrt{(l+m)!(l-m)!(l+m')!(l-m')!} \nonumber \\
&\times\sum_{k}\frac{(-1)^{k+m'-m}}{k!(l+m-k)!(l-m'-k)!(m'-m+k)!}  \nonumber \\
&\times (\sin{\frac{\beta}{2}})^{2k+m'-m}(\cos{\frac{\beta}{2}})^{2l-2k-m'+m}.
\end{align}
Due to the properties of the group $SO(3)$, the inverse transformation is then given by
\begin{equation}
\label{eq:A6}
Y_{lm}(\theta,\varphi)=\sum_{m'=-l}^l e^{-im'\gamma}d^l_{m'm}(-\beta)e^{-im\alpha}Y_{lm'}(\theta',\varphi').
\end{equation}
The next step is to include the change of spin-basis under a rotation. According to \cite{Alcubierre2008} a quantity $\eta$ of spin-weight $s$ obeys the following law under a change of the spin basis:
\begin{equation}
\label{eq:A7}
\eta'=\eta e^{is\chi}.
\end{equation}
Combining Eqs. (\ref{eq:A6}) and (\ref{eq:A7}) yields the transformation law for the spin-weighted spherical harmonics:
\begin{equation}
\label{eq:A8}
Y^s_{lm}(\theta,\varphi)=e^{-is\chi}\sum_{m'=-l}^l e^{-im'\gamma}d^l_{m'm}(-\beta)e^{-im\alpha}Y^s_{lm'}(\theta',\varphi').
\end{equation}
We invert Eq. (\ref{eq:A1}) to determine the transformation law for the $\Psi_{4,lm}$-modes,
\begin{align}
\label{}
 \Psi_{4,lm}&=\int\Psi_4\overline{Y^s_{lm}(\theta,\varphi)}d\Omega \nonumber  \\
  &=\int e^{-is\chi}\Psi'_4e^{is\chi}\sum_{m'}e^{im'\gamma}d^l_{m'm}(-\beta) \nonumber\\
  &\qquad \times e^{im\alpha}\overline{Y^s_{lm'}(\theta',\varphi')}d\Omega' \nonumber \\
  &=\sum_{m'=-l}^l e^{im'\gamma}d^l_{m'm}(-\beta)e^{im\alpha}\Psi'_{4,lm'},
\end{align}
where we see that explicit knowledge of $\chi$ as a function of $\theta$ and $\varphi$ is not necessary
to determine the coefficients $\Psi_{4,\ell m}$.
This transformation law can now be applied to any given $\Psi'_{4,lm}$, e.g., our numerical data, in order to change the frame of reference. The remaining free parameters are the three angles that determine the rotation. We restrict ourselves to a rotation about two Euler angles, $\beta$ and $\gamma$, only. Since we aim to align the orbital angular momentum with the $z$-axis at every instant of time, i.e., $\hat{L}\mapsto\hat{z}$, a simple calculation shows that in order to fulfill this $\beta=-\theta$ and $\gamma=-\varphi$ are required, where $(\theta,\varphi)$ are the polar coordinates determining the direction of $\hat{L}$.


\bibliography{OrbitalPlane}

\end{document}